\documentclass[a4paper,12pt]{extarticle}
\usepackage{geometry}
\usepackage[T1]{fontenc}
\usepackage{helvet}
\usepackage[utf8]{inputenc}
\usepackage{booktabs}
\usepackage{graphicx}
\usepackage{textcomp}
\usepackage[geometry]{ifsym}
\usepackage{float}
\usepackage{listings}
\usepackage{xcolor}
\usepackage{hyperref}
\usepackage{multirow}
\usepackage{subfig}
\usepackage[toc,page]{appendix}
\usepackage{pdfpages}

\begin{document}
\begin{titlepage}
    \begin{center}
        \vspace*{1cm}
            
        \Huge
        \textbf{Final Project: RNN-Transducer-based Losses for Speech Recognition on Noisy Targets}
            
        \vspace{0.5cm}
        \LARGE
            
        \vspace{1.5cm}
            
        \textbf{Vladimir Bataev}
            
        \vfill
            
        \vspace{0.8cm}

        \Large
        University of London\\
        10 March 2024
            
    \end{center}
\end{titlepage}

\tableofcontents
\listoffigures
\listoftables

\clearpage

\section{Project Concept and Motivation}

\subsection{Overview and Template}
In our project, we will train an automatic speech recognition (ASR) system on noisy targets. 
We start with the template "CM3015 Machine Learning and Neural Networks, Theme 1: Deep Learning on a public dataset," which describes the task of choosing a publicly available dataset and training a deep learning model on it. So, we will work with a neural network-based end-to-end ASR system, using LibriSpeech~\cite{panayotov2015librispeech} dataset, a popular academic benchmark. We limit our task to RNN-Transducer~\cite{graves2012transducer} systems, which are widely used in production and provide state-of-the-art quality~\cite{li2022recentsurvey} in most cases. 

We are going beyond the standard task and focusing our research on making RNN-Transducer systems robust to noisy targets: unlike well-curated datasets, in the industry, the training data contains different errors due to the unreliability of the transcription sources or the inability to transcribe noisy speech accurately. To solve the problem of training on the noisy data, we will analyze the impact of different types of errors in training data on the quality of the RNN-Transducer system and explore different loss modifications to overcome the problem. We will construct the artificial training data by mutating correct transcripts from the LibriSpeech~\cite{panayotov2015librispeech} training part, similar to the approaches used in the related work, and try to achieve the best possible quality on the development and test data standard for LibriSpeech.

\subsection{Motivation}

Training ASR systems usually requires a large amount of well-transcribed audio-text paired data. The process of dataset preparation often calls for filtering out "noisy" transcriptions using some pretrained model, which results in a lower amount of available data for training. It is hard to obtain large, well-transcribed datasets for many languages and scenarios. On the other hand, data with limited transcript quality is widely available. Developing new approaches for working with weakly supervised setups can be beneficial in the following ways: 
\begin{itemize}
    \item (1) making it easier to use non-well-curated datasets;
    \item (2) improving the quality of ASR models without tricky filtering pipelines for raw data, making an approach fully "end-to-end," and potentially getting benefits from more non-filtered data;
    \item (3) improving ASR quality for low-resource languages when the data is extremely limited;
    \item (4) transfer learning scenarios when another (imperfect) model transcribes the data.
\end{itemize}

\subsection{Related Work: Systems}

Most of the modern ASR systems use mel filter bank features extracted from the speech signal~\cite{graves2012transducer} and learn to map a sequence of feature vectors to the correct sequence of the units derived from the text (e.g., characters, subword units, words, or phonemes). 
There are three dominating types of end-to-end ASR systems~\cite{li2022recentsurvey}: Connectionist Temporal Classification (CTC) ~\cite{graves_connectionist_2006}, RNN-Transducer (RNN-T)~\cite{graves2012transducer} and attention-based encoder-decoder (AED)~\cite{chan2016las}. 
CTC and RNN-Transducer rely on an explicit latent monotonic alignment between the audio and corresponding transcript~\cite{prabhavalkar2023survey}. CTC is the most straightforward non-autoregressive system, which predicts each text unit independently. RNN-Transducer was introduced as a solution to fix the wholly conditional independence assumption of CTC and consists of 3 parts: an Encoder, which produces the representation of input features non-autoregressively; autoregressive Prediction network; and a Joint network, which combines their output and produces the final prediction~\cite{graves2012transducer}. Encoder-Decoder systems with Attention~\cite{chan2016las} implicitly learn the alignment between the audio and text via the attention mechanism. We are primarily interested in RNN-Transducers since such systems are suitable for streaming by design, widely used in production, and power most of the state-of-the-art monolingual models~\cite{prabhavalkar2023survey}. Initially, recurrent neural networks (RNNs) were used as an autoregressive prediction (thus, RNN-T is named after RNNs), but other non-recurrent architectures can also be used, e.g., a simple stateless network whose output depends on a fixed number of previous outputs~\cite{ghodsi2020statelessrnnt}.

\subsection{Related Work: Models}

In ASR, the dominating architecture for the encoder is a Conformer~\cite{gulati2020conformer, asrleaderboard,li2022recentsurvey}, which originates from a transformer block architecture~\cite{vaswani2017attentiontransformer} augmented with convolutional modules~\cite{gulati2020conformer}. The important part is that the original Conformer encoder subsamples the sequence of input features (derived from the audio signal) 4 times, and recently, multiple architectures were proposed to subsample the input features 8 times to provide better speed without performance drop, e.g., Fast Conformer~\cite{rekesh2023fastconformer}. We will use the Fast Conformer architecture as a basic architecture for our model, focusing on loss modifications rather than changing the architecture.

\subsection{Related Work: Losses}

\textbf{CTC} is the loss that can naturally be represented~\cite{laptev2021ctctopo} with weighted finite state transducers (WFSTs). Due to the existence of the libraries that allow to construct differentiable WFSTs and use them to train deep learning systems, e.g., k2~\cite{povey_k2}, different modifications of CTC loss were proposed to solve the problem of different errors in training data. Particularly, W-CTC (CTC with Wild Cards)~\cite{cai2022wctc} allows missing text at the start and the end of an utterance in transcription, Star Temporal Classification (STC)~\cite{pratap2022stc} allows missing labels anywhere, Bypass Temporal Classification (BTC)~\cite{gao2023bypassbtc} solves the problem of insertions and substitutions. Recently proposed Omni-temporal Classification (OTC)~\cite{gao2023omniotc} is a generalized loss that combines all the previous work and proposes a CTC loss modification that is robust to any type of errors in transcripts.

\textbf{RNN-Transducer} robustness to the errors in training data is still an unsolved problem. The recent work about a graph-based framework for RNN-Transducer~\cite{laptev2023rnntwfst} proposes a generalized solution to develop different loss modifications based on WFSTs and also proposes a W-Transducer loss that can deal with missing transcripts at the start and end of the transcription (similar to W-CTC~\cite{cai2022wctc}). Moreover, an autoregressive prediction network can require modifications since its input in training time is a ground truth transcription and can be sensitive to incorrect text. We are planning to increase the complexity of the project gradually, starting from the "under-transcribed" case, when the training texts can contain missing words (deletions) similar to STC~\cite{pratap2022stc}. Then, we will explore more complex cases with insertions and substitutions, finally providing a general combined solution.

\section{Literature Review}

\subsection{Introduction}

We start our review with the work on the CTC criterion and its modifications. Since both CTC and RNN-T take into account all possible alignments between features extracted from audio and text units, thus it is possible to apply some techniques to both of them. Then, we discuss the RNN-T loss and the relevant work to improve its robustness to errors, along with the structure of the Transducer models and differences between CTC and RNN-T that prevent direct application of CTC-based techniques.

\subsection{CTC and Modifications}
\subsubsection{Connectionist Temporal Classification (CTC) loss}
The Connectionist Temporal Classification (CTC) was originally introduced in~\cite{graves_connectionist_2006} as a replacement to a "classical" ASR pipeline based on hidden Markov Models (HMMs) and is historically the first so-called "end-to-end" ASR system. The original work proposes a solution for the automatic learning for the alignment between the target units (phonemes in this work, using the TIMIT~\cite{garofolo1993timit} dataset) and the representation extracted by the neural network from the audio signal. The system utilizes RNNs (particularly bidirectional Long Short-Term Memory (BLSTMs) networks~\cite{graves2005framewiseblstm}) as its backbone. Mel-Frequency Cepstrum Coefficients are extracted from the input audio every 10ms, which forms the input for the neural network. The output of the network has a softmax activation and is interpreted for each label as a probability of observing the label at the current time frame~\cite{graves_connectionist_2006}, as in the classification task for each frame. The vocabulary is augmented with a special $\langle blank \rangle$ symbol, and the algorithm considers all possible variants of transcriptions that map to the original text units after removing duplicated predictions and the $\langle blank \rangle$ symbol. The loss is a minus log probability of all possible correct alignments (defined by the rule described above) given the audio features. Thus, training maximizes the log probability for the possible alignments without requiring a forced alignment, which was used in classical HMM-based systems. The paper also proposes an efficient forward-backward algorithm to calculate the CTC loss and gradients and shows the system's efficiency in predicting phonemic transcriptions of utterances.

The paper about CTC loss~\cite{graves_connectionist_2006} forms a basis for our research since we plan to apply techniques to modify the possible alignments in the CTC framework to the RNN-Transducer system. The work shows that it is not necessary to have one perfect alignment as a target, but once we construct a mapping between the ground truth target and the inner latent alignment (in the paper - sequences with repeated labels and additional $\langle blank \rangle$ symbols, that are removed in decoding), we can use a full-sum training (taking into account all possible alignments), and the system can effectively solve the ASR task.

\subsubsection{Wild-card CTC (W-CTC)}
Wild-card CTC (W-CTC) was proposed in~\cite{cai2022wctc} to solve the problem when the utterance is partially transcribed, and the transcription can be missing at the start, at the end, or on both sides. The work introduces a simple but efficient modification of CTC criterion computation, using a special "*" (star) symbol that can be prepended to each transcription, and means that at the start of the transcription, any possible sequence (of any size, including the empty sequence) of symbols can be missing. This allows a simple modification of the dynamic programming algorithm for CTC computation introduced in~\cite{graves_connectionist_2006} to handle the alignments not only when the ground truth is entirely correct but also start and end the alignments between text units and features extracted from audio at any part of the audio, but with the assumption, that the part of the audio fully matches the transcription. The authors also use a TIMIT~\cite{garofolo1993timit} dataset, randomly masking a portion of the start/end of the utterance transcriptions, thus showing the effectiveness of the approach (compared to the original CTC) on partially transcribing ASR data in a synthetic setup. Additionally, the authors validate the effectiveness of W-CTC on Optical Character Recognition (OCR) and Continuous Sign Language Recognition (CSLR) tasks.

The paper is interesting as the first work which introduces a synthetic setup with under-transcribed data to study and develop training criteria for ASR robust to corrupted targets. The proposed W-CTC criterion uses a larger possible amount of alignments but still converges and surpasses CTC even when a small portion of the data is corrupted. Also, the paper shows the application of the techniques to OCR and CSLR tasks, which shows the potential impact of modifications of losses used for ASR on areas outside the speech recognition field.

\subsubsection{Star Temporal Classification (STC)}
Star Temporal Classification~\cite{pratap2022stc} was proposed as a generalization for the previous work when the transcription is only partial, and between any pair of labels, an arbitrary number of words can be missing. The work uses the "*" star token to represent zero or more text tokens and considers alignments between the encoded signal features and a target, where the "*" token is inserted between all labels and also appended to the start and the end of the utterance. Thus, the loss allows the network to output any sequence of labels corresponding to the presented corrupted ground truth text after removing some of the words (the ASR system should emit all the words from the target but can insert other words between them). This approach significantly increases the number of possible alignments and does not allow the training of the neural network directly. The problem is solved by introducing a penalty for the "*" token $\lambda$, with exponential decay during training, starting from the high values and decreasing once the network converges. The penalty hyperparameter is also adjusted based on the number of missing words in transcriptions. The authors demonstrate the approach using synthetic data derived from LibriSpeech~\cite{panayotov2015librispeech}, using partially masked transcripts with different probabilities (up to 70\%). The approach is practical even when using greedy decoding, but the authors also show that decoding with an n-gram language model (LM) and rescoring hypotheses with Transformer~\cite{vaswani2017attentiontransformer} based LM also improves the system's quality. For the implementation, unlike previous work, the GTN~\cite{hannun2020Differentiablegtn} framework for differentiable WFSTs is used, which simplifies the development of the new loss.

This paper is crucial for our work since it allows us to solve the problem with deletions in the transcribed texts for CTC. The critical part is that the direct solution (considering all possible alignments in the loss) leads to failure, and using the adjustable penalty is crucial to solve the issue. Also, the construction of LibriSpeech-based data is an important part. We will consider an analogous approach for investigating similar cases for RNN-T.

\subsubsection{Bypass Temporal Classification (BTC)}
Bypass Temporal Classification~\cite{gao2023bypassbtc} solves part of a weakly-supervised setup with unreliable transcripts: substitutions and insertions. Similar to the previous work, the authors propose CTC target graph modification to allow the alignments when some text units from the target are changed (substitutions) or not used (insertions in the text, thus allowing deletion from the alignment). Remarkably, the ground truth is represented as a linear WFST, where the forward arcs contain ground truth labels, and each arc also presents a parallel "bypass" arc with a penalty, which allows to skip the token by emitting zero or more tokens. The authors also use the penalty with an exponential decay to make the network learn using the increased number of possible alignments. The authors evaluate their solution on both TIMIT~\cite{garofolo1993timit} and LibriSpeech~\cite{panayotov2015librispeech} datasets, constructing the training data with substitutions and insertions separately and combining them. Interestingly, it is shown that for the CTC criterion, the impact of insertions is larger than for substitutions, and with 50\% of insertions, it is impossible to train a system with the pure CTC criterion.

This paper solves the remaining piece of the puzzle of training a CTC-based system with partially incorrect transcripts. We will also investigate similar approaches for RNN-T. Moreover, this paper encouraged us to start the exploration of the problem by comparing the impact of different errors on the training criterion behavior and starting our solution from the most disruptive problem.

\subsubsection{Omni-temporal Classification (OTC)}
The Omni-temporal Classification~\cite{gao2023omniotc} finally combines the approaches to construct a universal loss to handle all the possible cases of errors. The authors utilize the "bypass" arcs to skip frames, along with self-loops that represent substitutions and insertions, and two separate penalties for them following the previous work. The authors used synthetic data generated from the train-clean-100 subset of LibriSpeech~\cite{panayotov2015librispeech} and also trained a system on the original subsets from LibriVox~\cite{librivox} that were originally used to construct the LibriSpeech data, using segmentation and filtering with pretrained models, as described in~\cite{panayotov2015librispeech}. The combination of the losses leads to significant improvements for all types of errors and allows training of the ASR system even on the original unsegmented data with errors without a cleaning pipeline.

This work contains a detailed exploration of the impacts of different penalty values on the quality of the network, which can provide some insights for working with similar alignment-based modification approaches for RNN-T. The code is published, and we can try to reproduce the setup if necessary. Moreover, we can try this criterion as a part of the potential solution with hybrid CTC-RNN-T models for a "CTC head," which we will discuss further. Moreover, we found it curious that despite the beneficial impact of the loss when training ASR system with unfiltered raw data from LibriVox, the gap still exists for the carefully filtered LibriSpeech data (e.g., authors report~\cite{gao2023omniotc} 12.5\% vs 8.2\% WER on test-other for these setups), which can indicate that the problem is still not fully solved even for CTC criterion, and multi-stage pipeline is still essential to obtain the best performance.

\subsection{RNN-Transducer and Modifications}
\subsubsection{RNN-T loss}
Recurrent Neural Network Transducer (RNN-Transducer, RNN-T)~\cite{graves2012transducer} was developed as an improvement of the CTC~\cite{graves_connectionist_2006} systems to overcome problems related to conditional independent assumption. The system contains three components: (1) transcription network, which is usually referred to as an "Encoder"(e.g., ~\cite{He2018StreamingES}), that operates on the features derived from audio and is similar to the one used in CTC system; (2) prediction network, that outputs the predictions based on the previously decoded symbol in inference time, and utilizes a ground truth text in training time; and (3) a combination of the outputs of the previous two networks that makes final predictions, which is usually referred as a Joint network in the literature~\cite{He2018StreamingES}. The prediction network is autoregressive and provides a crucial improvement for the ASR system. The system also uses a $\langle blank \rangle$ symbol, but the meaning differs from CTC: this symbol means the system should end decoding the current frame and transition to the next frame from the encoder. Unlike CTC, repeated symbols are not allowed, but the advantage is that the system can predict more than one text unit for each frame. The RNN-T loss function takes into account all possible monotonic alignments between encoder and prediction network outputs, which makes the system similar to CTC with implicit alignment learning.

The crucial difference for the RNN-T system is that the prediction network is autoregressive (usually uses LSTM as a backbone), making it significantly more challenging to deal with imperfect transcripts. Since ground truth text is used during training (so-called "teacher forcing" algorithm), corrupted tokens can prevent the correct alignment learning. The autoregressive nature also prevents applying CTC-based approaches directly to RNN-T and may require modifications of the network itself along with the training criterion.

\subsubsection{Graph-based RNN-Transducer framework}
The recent work about RNN-Transducer modifications~\cite{laptev2023rnntwfst}\footnote{\begin{scriptsize}Disclaimer: the author of the Final Project participated in the development of this approach and is a co-author of the paper.\end{scriptsize}} brings the connection between WFSTs and Transducer architecture, allowing representing the RNN-T loss computation with the graph, where the arc weights are taken from the Joint network output, and thus the direct application of graph algorithms, including full-sum alignment learning, are possible. The work presents two approaches for representing original and modified RNN-T graphs, either as a direct grid ("Grid-Transducer") or as a composition of acoustic and textual schemas ("Compose-Transducer"), which after the composition and connect operations exactly matches the Grid-Transducer representation ("connect" operation removes the states that do not belong to any path from the start and the end state; it is optional for loss computation since such states do not affect the alignment probabilities since not belonging to any alignment). The composition is slower to compute but allows for the development of losses using graphs that can be easily debugged visually. The authors build the framework on top of k2~\cite{povey_k2} library for differentiable WFSTs and show that the loss computation can be as efficient as the optimized CUDA-based code. Moreover, the authors present a W-Transducer, which can solve the problem of deletions at the start and the end of the utterance, similar to the W-CTC discussed above. The graph for the loss uses two groups of skip-connections from the start of the time grid to all other time steps and from each time step in the grid to the last, which allows to use the alignments where some frames at the start and the end are skipped, but the network should emit the full training text. The effectiveness of this approach was demonstrated on LibriSpeech data by randomly removing 20\% and 50\% of the labels from text from both sides of each utterance.

In our work, we will use the proposed WFST framework for training RNN-T on noisy targets to simplify the development of the losses. W-Transducer solves the easiest case when the autoregressive prediction network does not use a corrupted input and thus can remain unmodified. We are planning to work with more complex cases, and such modifications can be required along with the loss customization. We will discuss this approach in more detail in Section~\ref{sec:design-rnnt-explained}.

\subsubsection{Stateless RNN-T}
The work~\cite{ghodsi2020statelessrnnt} proposes "RNN-T with StateLess Prediction Network (RNNT-SLP)," revising the need for recurrent networks for the prediction network part of the RNN-T system. The authors questioned the popular opinion that the prediction network behaves similarly to the language model in traditional ASR systems. They tried pretraining the recurrent network on text-only data to predict the next symbol and initializing the prediction network with pretrained weights and found that such pretraining does not lead to any improvement. Moreover, replacing the prediction network with a simple "stateless" module, in which prediction is based only on the last symbol (which is similar to 2-gram LM), leads to a comparable overall system performance with the RNN-based system. 

Despite there are other works that show that the prediction network can behave like a language model, especially in a factorized architecture~\cite{variani2020hat,meng2022mhat}, we are interested in this work showing the possibility of using simpler networks for this part of the model. Since a stateless prediction network can be significantly more robust for corrupted targets, in which prediction is dependent only on a small context, we consider this work an important option for changing the prediction network architecture.

\subsubsection{Hybrid CTC-Transducer models}
As a last item of our review, we will discuss the hybrid architecture, which uses both RNN-T and CTC losses, proposed in~\cite{tian2021fsrhybrid}. The work proposes training the neural transducer with an auxiliary CTC loss on top of the encoder (as a separate "head"), combining the systems for further accelerating the decoding: if the CTC head predicts the blank label, such frame can be skipped in decoding, resulting in a significant inference acceleration with tiny quality degradation. The work ~\cite{wang2022hybridctcrnnt} proposes the use of such hybrid systems to accelerate not only inference but also the training speed, using more lightweight CTC head prediction to compute RNN-T loss with the smaller number of possible alignments (pruning RNN-T loss lattice).

This work can be relevant for our research due to the discussed above OTC criterion, which is robust to noisy targets: in a hybrid training setup, we can use the prediction of the additional head, trained with the OTC loss, to identify corrupted tokens and thus modify the input for prediction network or the loss based on this information.

\section{Project Design}

In our project, we plan to investigate the behavior of the RNN-Transducer architecture in a weakly supervised setup when part of the data is corrupted and modify the RNN-T loss to improve the system's quality. We will consider loss modification techniques that do not require any change in the overall neural architecture and decoding algorithms to make them compatible with current production systems.

\subsection{Objectives}

We define the following objectives for our work:
\begin{itemize}
    \item (1) Investigate how using partially incorrect transcripts impacts the quality of the RNN-Transducer system, separating deletions, substitutions, and insertion cases
    \item (2) Investigate techniques that allow to improve the performance of the system in the conditions described above
    \item (3) Combine the solutions provided in (2) to solve the general case of training the ASR Transducer-based system with unreliable transcript and evaluate the final solution when it is unknown what type of errors the data contains.
\end{itemize}

\subsection{Metrics}
\label{sec:design-metrics}

The key metric for assessing the quality of an ASR system is word error rate (WER)~\cite{wiki:wer}. The additional metric commonly used to assess the speed of the ASR system is a real-time factor (RTF), which indicates how much audio the system can process given the fixed time, and usually, the tradeoff between speed and quality is important when considering different models. In our project, we focus only on modifications that have a minor impact on the inference speed after the model is trained; thus, we report only WER in our experiment.

The word error rate is a specification of a Levenshtein distance~\cite{wiki:wer,wiki:editdist}, defined for two sequences of words (hypothesis and ground truth), and is calculated as a number of substitutions (SUB), insertions (INS) and deletions (DEL) divided by the number of ground truth (correct) words.
$$WER = \frac{SUB + INS + DEL}{CORRECT}$$
Lower WER means that the system makes more accurate predictions. We will also use the term "accuracy," usually defined as $accuracy = 1 - WER$ (higher accuracy value means better prediction).

Since we are working with the conditions that show the degradation of the standard ASR training pipeline, we introduce two additional metrics. WER difference (WERD) indicates the system degradation and is a difference between the WER that the system achieves in a particular setup and the WER of the baseline on the non-modified (original) data. $$WERD = WER_{modified\_data} - WER_{original\_data}$$ We are interested in minimizing the degradation of the system, thus defining the relative improvement of the system as WERDR. $$WERDR = \frac{WERD_{RNN-T} - WERD_{Proposed}}{WERD_{RNN-T}}$$

\subsection{Data}

The primary data for the project is a LibriSpeech~\cite{panayotov2015librispeech} corpus, which consists of 3 subsets for training data (960 hours total), two development sets (\textit{dev-clean} and \textit{dev-other}, 5.4 and 5.3 hours respectively), and two test sets (\textit{test-clean} and \textit{test-other}, 5.4 and 5.1 hours). We will use the full training part to generate artificial training data by corrupting the texts with artificial deletions, substitutions, and insertions. We will use the \textit{dev-other} set for validation during training and choosing the optimal checkpoints, and we will finally assess our models on the \textit{test-other}. As it is common in the ASR research, we will report WER for all development and test sets.

\subsection{Backbone Encoder Models}

Currently, the dominating architecture~\cite{prabhavalkar2023survey} for the encoder is the Conformer~\cite{gulati2020conformer} model. The original Conformer-encoder processes the input sequence of vectors and produces the representation 4 times smaller by the time dimension (4x subsampling). For our initial experiments, we will use Fast Conformer~\cite{rekesh2023fastconformer} (114M parameters), which subsamples the input by the factor of 8 without accuracy degradation by using depth-wise separable convolutions~\cite{Chollet2016depthwisesep}, which allows faster training and inference.

\subsection{Tools and Frameworks}

We are planning to use NeMo~\cite{kuchaiev2019nemo} framework for experiments, which is based on PyTorch-Lightning~\cite{Falcon_PyTorch_Lightning_2019} for easiness of extending the models, and train them on clusters. NeMo provides stubs for ASR models and capabilities to use WandB~\cite{wandb} platform for experiment management. When required, we will use pure PyTorch~\cite{paszke2019pytorch} and k2~\cite{povey_k2} library to modify losses and prediction network, also using a WFST framework for RNN-Transducer~\cite{laptev2023rnntwfst} implemented in NeMo.

\subsection{RNN-Transducer Details and Graph-based representation}
\label{sec:design-rnnt-explained}

\begin{figure}[t]
\centering
\includegraphics[width=16cm]{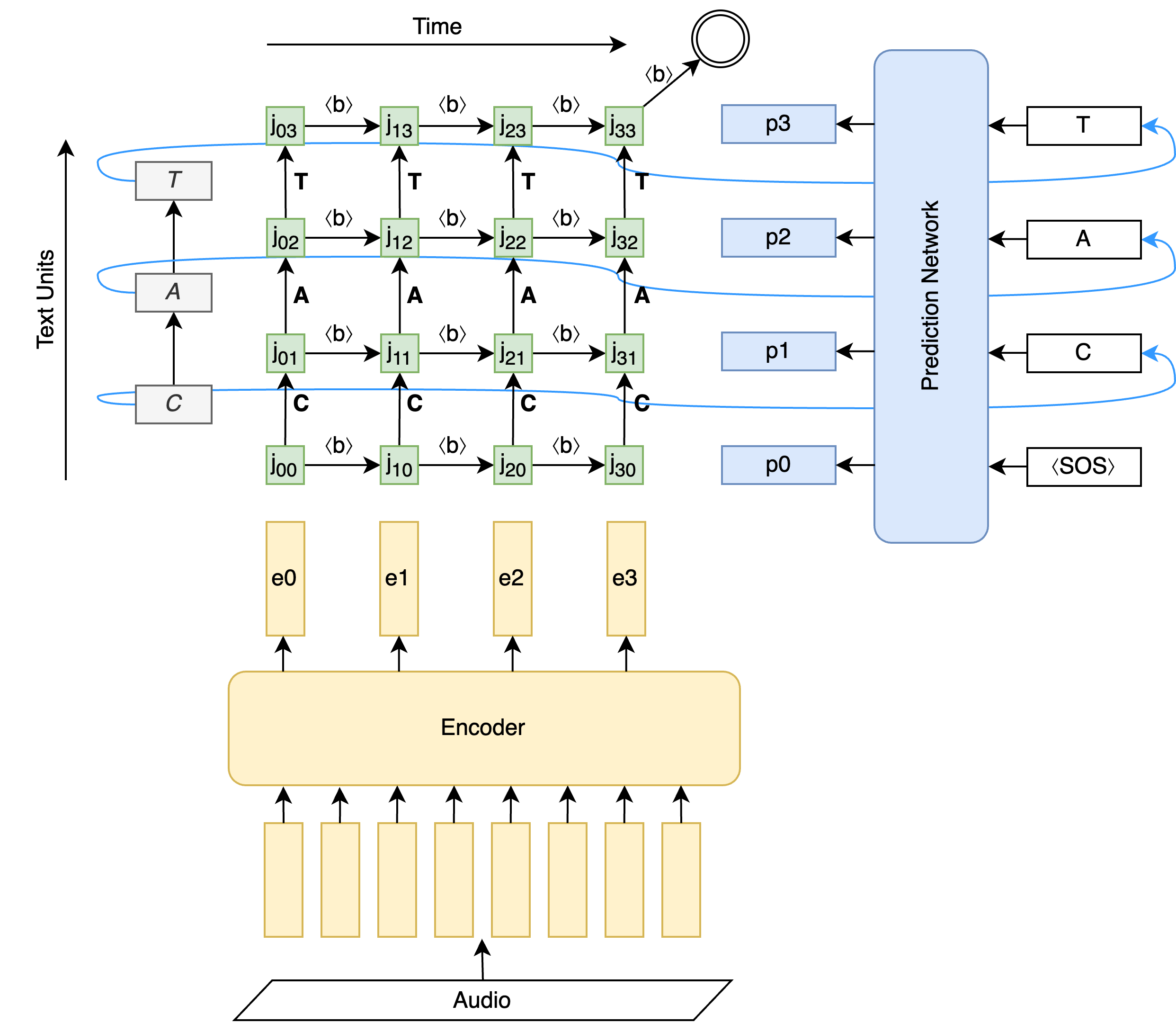}
\caption{RNN-Transducer Schema}
\label{fig:rnntransducer}
\end{figure}

RNN-Trasducer~\cite{graves2012transducer} model schema is shown in Fig.~\ref{fig:rnntransducer}. The model consists of three parts. \textbf{Encoder} neural network transforms a sequence of input feature vectors derived from an audio into a latent representation - a sequence of vectors $\langle e_0,e_1,...,e_{t-1} \rangle$ with the length of $T$. The number of output vectors is usually smaller than the input due to applying strided convolutions or pooling layers along the time axis, which reduces the output size by a fixed factor (subsampling) and makes the model more computationally efficient. The \textbf{Prediction network} (usually a recurrent network) in training time takes a ground truth sequence of text units as an input padded with a special start-of-sequence symbol $\langle SOS \rangle$ (in practice usually $\langle blank \rangle$ symbol is reused for this purpose) and produces a latent representation – sequence of vectors $\langle p_0, p_1, ..., p_{u} \rangle$ with the length of $U+1$ ($U$ is the number of text units in the text representation). The \textbf{Joint network} combines all combinations of $e_i$ with $p_j$ vectors and produces a 3-dimensional tensor for the size $Tx(U+1)xV$, where $V$ is a vocabulary size augmented with the $\langle b \rangle$ ($\langle blank \rangle$) symbol (in practice, the 4-dimensional tensor is used due to additional batch dimension). Thus, $j_{i,j}$ represents the vector produced from a combination of $p_i$ and $p_j$. In practice, the Joint network is relatively simple and computes the sum of vectors, non-linearity (e.g., ReLU), and a projection into the space of size $V$, (e.g., $j_{i,j} = Project(ReLU(e_i + p_i))$). If the vectors $e_i$ end $p_j$ are of different sizes, they are firstly projected to the space with the same dimension. The Softmax activation produces a distribution over the vocabulary units, and the dynamic programming algorithm is used to compute the probability over all possible paths in the graph (in the code, all computations are produced in log scale for numerical stability purposes). Each path represents a possible alignment, where blank symbols can be inserted between labels (e.g., for the Fig.1~\ref{fig:rnntransducer} "C $\langle b \rangle$ A T $\langle b \rangle$ $\langle b \rangle$ $\langle b \rangle$" is the possible path for the target text "CAT" represented as "C," "A," and "T" units). The number of $\langle b \rangle$ labels is the number of frames the Encoder produces. The minus log probability of all possible alignments forms an RNN-T loss value. 

For \textbf{greedy decoding}, since the ground truth text is unknown, a nested loop is used: for each encoder vector, the algorithm sequentially obtains the next prediction with the maximum probability, starting from $\langle SOS \rangle$ symbol, and computes the Prediction network output for the new decoded symbol, combining with the current encoder vector and computing Joint network output. Once the $\langle b \rangle$ symbol is found, this symbol is not fed into the Prediction network, but the inner loop stops, and the decoding process starts for the next encoder vector. More complex decoding algorithms exist, but greedy decoding is widely used in practice due to the best speed and near-optimal quality in most scenarios~\cite{prabhavalkar2023survey}, so we are focusing on this approach.

\begin{figure}[ht]
    \centering
    \subfloat[\centering RNN-Transducer Unit Schema. Labels: $(text\_unit,unit\_position):text\_unit$]{{
        \includegraphics[width=7.5cm]{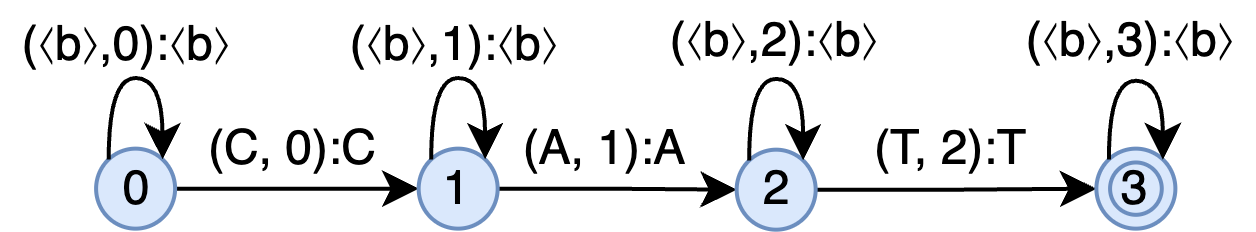}
        \label{fig:rnntunit}
    }}
    \subfloat[\centering RNN-Transducer Time Schema. Labels: $text\_unit:frame\_number$]{{
        \includegraphics[width=7.5cm]{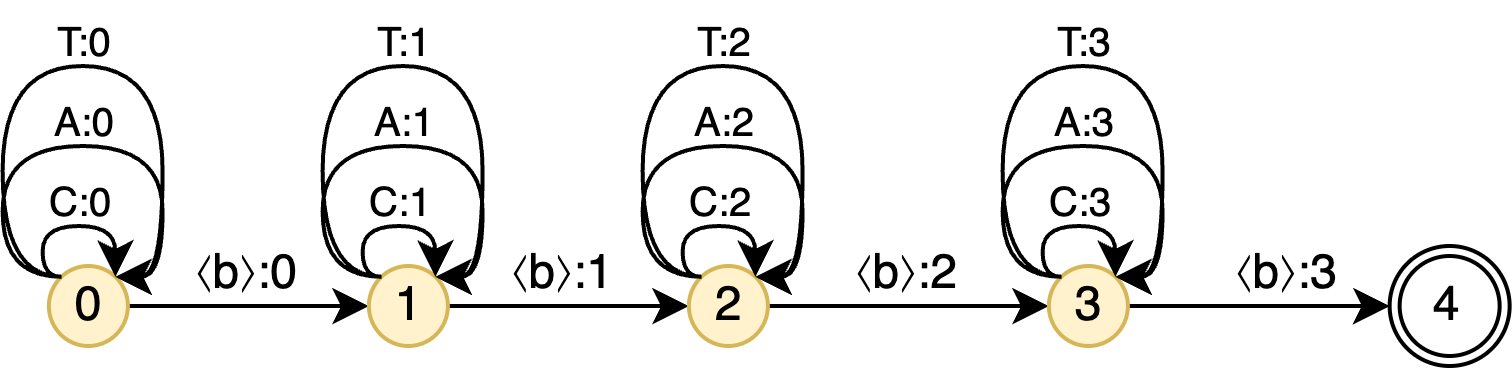}
        \label{fig:rnnttime}
    }}
    \qquad
    \subfloat[\centering RNN-Transducer Lattice. Labels: $(text\_unit,unit\_position):frame\_number$]{{
        \includegraphics[width=15cm]{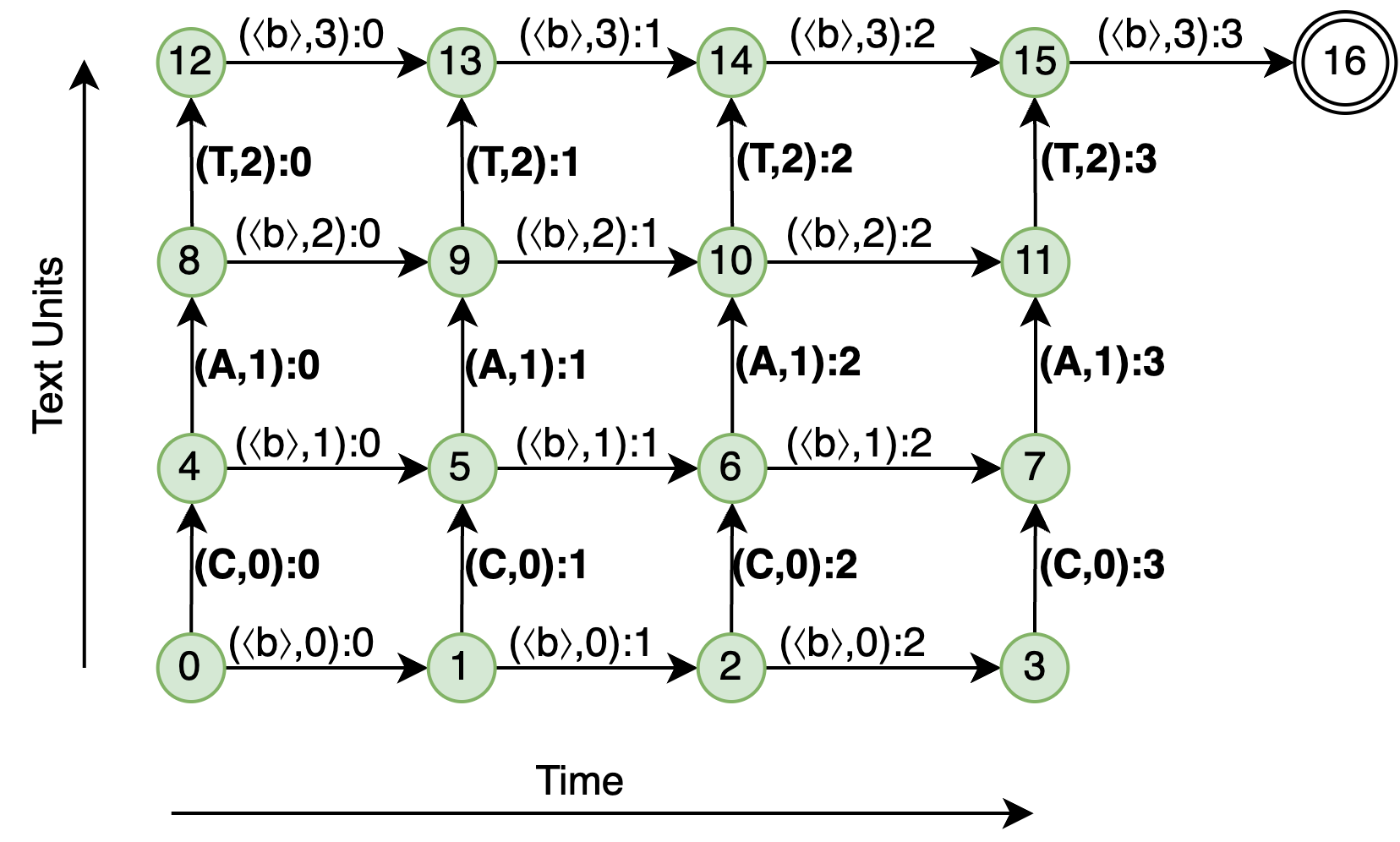}
        \label{fig:rnntlattice}
    }}
    \caption{WFSTs for RNN-Transducer, following~\cite{laptev2023rnntwfst}}
    \label{fig:rnntschemas}
\end{figure}

The computation of the RNN-T loss can be represented with \textbf{WFSTs}, as shown in~\cite{laptev2023rnntwfst}. The original work presents the approach more theoretically, and here we will discuss the practical implementation, which is a basis for our work. A weighted finite state transducer is a graph with states and arcs, whose arcs represent transitions from input to output labels with a corresponding weight. In k2~\cite{povey_k2} library, it is allowed to store an arbitrary number of labels (not only an input/output label for each arc). Thus, we use a tuple of labels as input labels for some graphs in our representation~\ref{fig:rnntschemas}. We can construct a computational graph using 3 labels for each transition: output (ground truth) label, unit index (in a sequence of ground truth units), and encoder vector index, see~\ref{fig:rnntlattice}. Given these labels by using "index select" \footnote{\begin{scriptsize}\url{https://pytorch.org/docs/stable/generated/torch.index_select.html}\end{scriptsize}} operation, we can populate a lattice with the weights from the Joint network output, and apply a differentiable computation of the full-sum loss\footnote{\begin{scriptsize}\url{https://k2-fsa.github.io/k2/python_api/api.html?highlight=get_tot_scores\#k2.Fsa.get_tot_scores}\end{scriptsize}}. This is a "Grid-Transducer," described in~\cite{laptev2023rnntwfst}. To simplify the development, we can construct two schemas ("Compose-Transducer"). Unit schema~\ref{fig:rnntunit} represents text units, and each transition has a tuple of input symbols $(Unit:Unit\_Index)$ and an output symbol $Unit$. $\langle b \rangle$ units represent self-loops, and other units represent transitions over the ground truth text. The time schema~\ref{fig:rnnttime} is a simple linear graph representing transition over time and self-loops for each state with all the vocabulary symbols as inputs and time index as output. It is much easier to visually debug the representation with the separated schemas, and their composition matches the final lattice~\ref{fig:rnntlattice}. We are planning to experiment with the separated schemas to make the RNN-T loss more robust to corrupted targets, but for the final stage, we are planning to provide an efficient code for lattice construction.

\subsection{Plan}
\label{sec:design-plan}

\subsubsection{Initial Project Plan}
\begin{figure}[t]
\centering
\includegraphics[width=16cm]{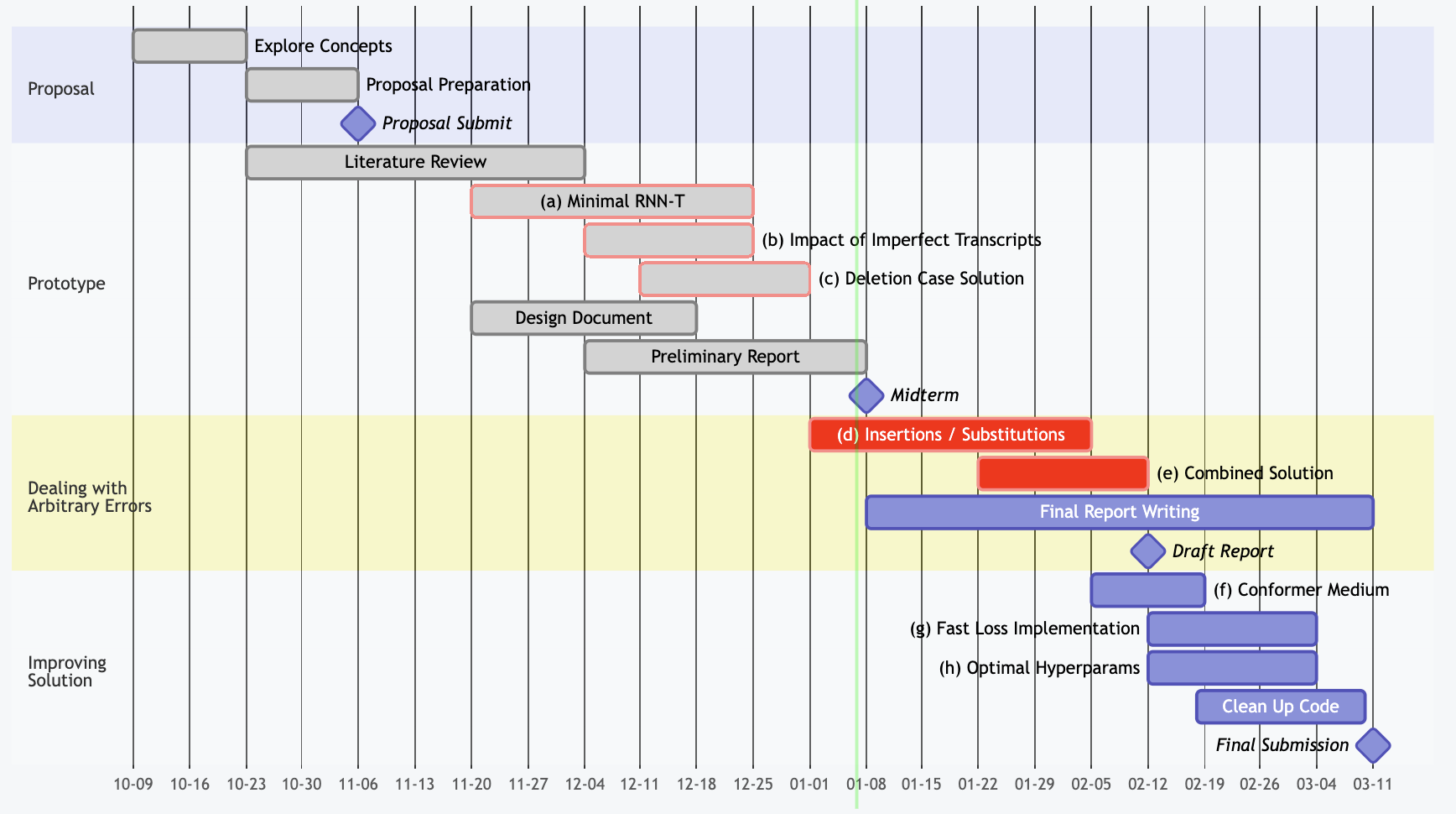}
\caption{Project Plan (Gantt Chart)}
\label{fig:gantt}
\end{figure}

Our work initial work plan published in the "Draft Report" (Midterm) is shown in Fig.~\ref{fig:gantt}. 

\textbf{(a) Minimal RNN-T Implementation}. After initial preparation, which was already done before finishing this report (initial exploration, project proposal, literature review), we will implement a minimal codebase to experiment with RNN-T. During our exploration, we found that RNN-T models in NeMo provide a lot of functionality but are not easy to extend. So, we will implement the lightweight pipeline with the following capabilities:
\begin{itemize}
    \item Lightweight Joint and Prediction networks
    \item Greedy Decoding
    \item Full model training pipeline
    \item Compatibility with NeMo encoders (including Conformer-based encoders as described above), since we are not planning to customize the encoder part.
\end{itemize}

Then, we plan to train the baseline on the original LibriSpeech data with a Fast Conformer encoder to check that our model can achieve comparable quality with the native implementation of RNN-T in NeMo.

\textbf{(b) Impact of imperfect transcripts}. We will generate the data derived from LibriSpeech with 20\% and 50\% of deletions, substitutions, and insertions separately (6 sets in total) and train the models to study the model's behavior in such conditions.

\textbf{(c) "Deletions" case.} Our preliminary studies showed that the deletions in the transcripts are the hardest case for RNN-T. We will try to solve this case by modifying the loss function. At this stage, we focus on improving the performance and not trying to find the perfect hyperparameters and/or provide the fastest implementation for the loss function.

\textbf{(d) "Insertions" and "substitutions."} We will explore different approaches discussed in the literature review to improve the system's performance in such conditions.

\textbf{(e) Combined Solution}. Combining the solutions (c) and (d) will allow for solving the universal problem of partially correct transcripts. We will generate additional training data for cases with all types of errors.

\textbf{(f) Conformer Medium (4x subsampling).} At this stage, we will apply our solution to the Conformer Medium model to study its behavior when the encoder reduces the input 4 times.

\textbf{(g) Fast loss implementation.} This task focuses on the quality and speed of our code. We will provide a clean and fast "ready to use" solution for our loss and prediction network modifications.

\textbf{(h) Study hyperparameters in detail.} This task will allow us to get insights about the hyperparameters of the systems proposed in (c)-(e). 

\subsubsection{Project Plan Reflection}

The sections (a)-(b) were crucial for our project but imposed minimal risks due to well-explored existing solutions. The most critical risks came from (c) and (d) cases since, according to our knowledge, no solutions existed for such tasks. To mitigate these risks, we also considered using a hybrid CTC-Transducer architecture with OTC loss, which can solve the problem, at least for the CTC head, and we can use its predictions to train the Transducer part instead of corrupted ground truth labels. Combined solution (e) was a key part of finalizing our project. 
We considered parts (f)-(h) as a fair improvement but not an essential contribution to our work and planned to focus on them only after finishing the main part.

In our final stage, we elaborated a successful solution for all discussed cases, providing a "drop-in" replacement for the RNN-Transducer loss for the cases without changing the model architecture at all. We also provided a fast implementation for the losses. Due to a lack of computational resources and difficulties training the system within the novel setup, we focused more on exploring parameters for Bypass Transducer loss, as discussed in Section~\ref{sec:bypass-transducer-eval}. We omitted experiments with Conformer Medium (f) and left them for future work.

\section{Implementation}

The implementation is published in the \textbf{GitHub repository}: \url{https://github.com/artbataev/uol_final}.
In this section, we describe the implementation with the links to the original files. Additional visualization of the produced lattices for all proposed losses can be found in the Jupyter Notebook \url{https://github.com/artbataev/uol_final/blob/main/notebooks/Loss_Demo.ipynb}.

\subsection{Project Implementation Overview}

The main goal of the project is to provide a solution to deal with different types of errors in the RNN-Transducer framework. We want to make our models comparable and compatible with the publicly available state-of-the-art models and want further to propose solutions to the NeMo~\cite{kuchaiev2019nemo} framework. So, we reuse components from NeMo, focusing on customization and modifications of the necessary parts. 

Firstly, we make a minimal necessary code of RNN-T for our experiments, providing implementation containing the RNN-T model and customizable Joint and Prediction networks. We reuse Conformer~\cite{gulati2020conformer} blocks from NeMo~\cite{kuchaiev2019nemo} for the Encoder network.

The Prediction network is a 1-layer LSTM with 640 hidden units, implemented \\ in \href{https://github.com/artbataev/uol_final/blob/main/min_rnnt/modules.py#L88}{\texttt{MinPredictionNetwork}} class. The Joint network, as discussed in Section~\ref{sec:design-rnnt-explained}, applies two projections of the output of Encoder and Prediction networks (2 linear layers) to the dimension of 640, sums the vectors, applies ReLU non-linearity, and projects the output (one more linear layer) into 1025-dimensional space (1024 BPE units and $\langle b \rangle$). It is implemented in \href{https://github.com/artbataev/uol_final/blob/main/min_rnnt/modules.py#L53}{\texttt{MinJoint}} class.

We also implement a greedy decoding algorithm for evaluation in \href{https://github.com/artbataev/uol_final/blob/main/min_rnnt/decoding.py}{\texttt{min\_rnnt/decoding.py}}. 

In our experimental setup, we are following the Fast Conformer~\cite{rekesh2023fastconformer} training pipeline\footnote{\begin{scriptsize}\href{https://github.com/NVIDIA/NeMo/blob/v1.21.0/examples/asr/conf/fastconformer/fast-conformer_transducer_bpe.yaml}{github.com/NVIDIA/NeMo/blob/v1.21.0/examples/asr/conf/fastconformer/fast-conformer\_transducer\_bpe.yaml}\end{scriptsize}}. The encoder has 108.7M parameters, prediction network 3.9M, and Joint 1.4M (totally 114M). 

\subsection{Data Preprocessing}

We preprocess LibriSpeech~\cite{panayotov2015librispeech} data and apply speed perturbation with rates $0.9$ and $1.1$ (3x audio data), using the published preprocessing script\footnote{\begin{scriptsize}\url{https://github.com/NVIDIA/NeMo/blob/v1.21.0/scripts/dataset_processing/get_librispeech_data.py}\end{scriptsize}} to make our pipeline comparable with published models. 

We use log-mel filterbanks extracted from audio every 10ms with the window 25ms and apply SpecAugment~\cite{park2019specaugment} in training. We use the vocabulary of 1024 BPE~\cite{bpe} tokens extracted using SentencePiece~\cite{kudo2018sentencepiece} library for text units. 

\subsection{Model and Training Pipeline}

We set up training of our model for 200 epochs using AdamW~\cite{loshchilov2017decoupledadamw} optimizer with Cosine annealing~\cite{loshchilov2016sgdrcos} learning rate schedule with a linear warmup for 40 epochs and the maximum learning rate of $5e-3$. For experiments except for the baseline, we stopped training the model after 60 epochs since we are interested in the relative difference in model quality, and achieving the best possible accuracy is not our priority at this stage. We are reporting the results for the best checkpoint chosen on the \textit{dev-other} validation set.
For all experiments, we maintain a global batch size of 2048. We are training models on clusters using NVIDIA A100 (mixed-precision with bfloat16) and V100 GPUs (float32 full-precision), and depending on the availability of the resources varying local batch size from 8 to 32 to fit into memory and adjusting gradient accumulation to make the global batch size constant. We did not observe any difference in quality for a fixed global batch size when using an arbitrary number of nodes, varying local batch size, and using mixed or full precision. So, we are not reporting these details for each experiment.

\subsection{Proposed Losses}

In our work, we propose three modifications of the RNN-T loss:
\begin{itemize}
    \item Star-Transducer to dial with arbitrary deletions
    \item Bypass-Transducer to solve the case of insertions
    \item Target-Robust-Transducer, which is the combination of the previous modifications, allows to mitigate the problems of substitutions in target texts and also can be used as a universal loss when the type of errors is unknown.
\end{itemize}

The modifications are implemented in \href{https://github.com/artbataev/uol_final/tree/main/min_rnnt/losses}{\texttt{minrnnt/losses}} subpackage. All the classes follow Graph-RNNT~\cite{laptev2023rnntwfst} framework and inherit \texttt{GraphRnntLoss} class from the NeMo~\cite{kuchaiev2019nemo} framework and reuse its methods. The implementation uses k2~\cite{povey_k2} library. As described in Section~\ref{sec:design-rnnt-explained}, the computational lattice can be constructed 
as a composition of temporal and unit schemas, implemented in \texttt{get\_temporal\_schema} and \texttt{get\_unit\_schema} respectively. The faster implementation constructs the lattice directly in \texttt{get\_grid} method. In the initial development, we used the composition and then made the \texttt{get\_grid} implementation as a faster option. We also customize the \texttt{forward} method to assign appropriate scores to the arcs corresponding to special tokens, as described below.

For all losses, we add unit tests (in \href{https://github.com/artbataev/uol_final/tree/main/tests}{\texttt{tests}} directory) to make the sanity check for the following:
\begin{itemize}
    \item Graphs produced by composition of the temporal and unit schemas are equivalent to the graph produced by \texttt{get\_grid} method.
    \item When the weight of special arcs is $-\infty$, this is the equivalent of removing such arcs from a computational graph ($e^{-\infty} = 0$, such a transition does not contribute to loss computation); and the loss should be equivalent to original RNN-T loss. This is tested by comparing the loss value and gradient based on random input for the proposed loss and etalon RNN-T implementation.
\end{itemize}

The graph construction is debugged visually in the Jupyter Notebook~\cite{Kluyver2016jupyter} using automatic visualization from the k2~\cite{povey_k2} library with GraphViz package~\cite{graphviz}.

\subsubsection{Star-Transducer (Star-T)}
\label{sec:star-transducer-explained}

\begin{figure}[ht]
    \centering
    \subfloat[\centering Star-Transducer Unit Schema. Labels: $(text\_unit,unit\_position):text\_unit$]{{
        \includegraphics[width=7.5cm]{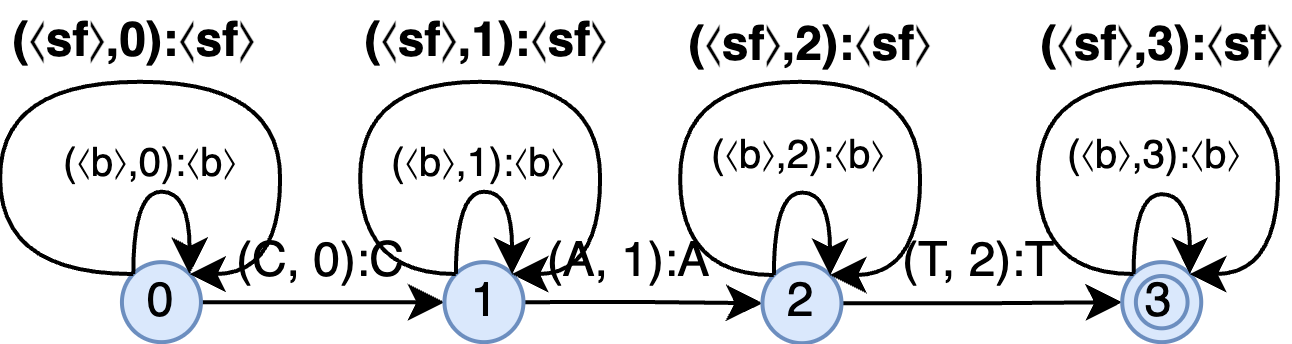}
        \label{fig:startunit}
    }}
    \subfloat[\centering Star-Transducer Time Schema. Labels: $text\_unit:frame\_number$]{{
        \includegraphics[width=7.5cm]{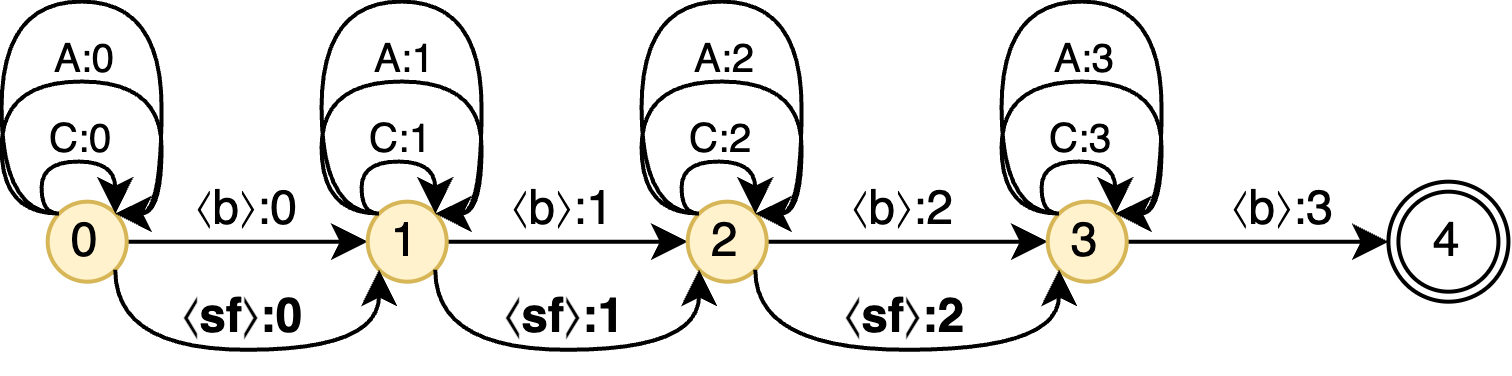}
        \label{fig:starttime}
    }}
    \qquad
    \subfloat[\centering Star-Transducer Lattice. Labels: $(text\_unit,unit\_position):frame\_number$]{{
        \includegraphics[width=12cm]{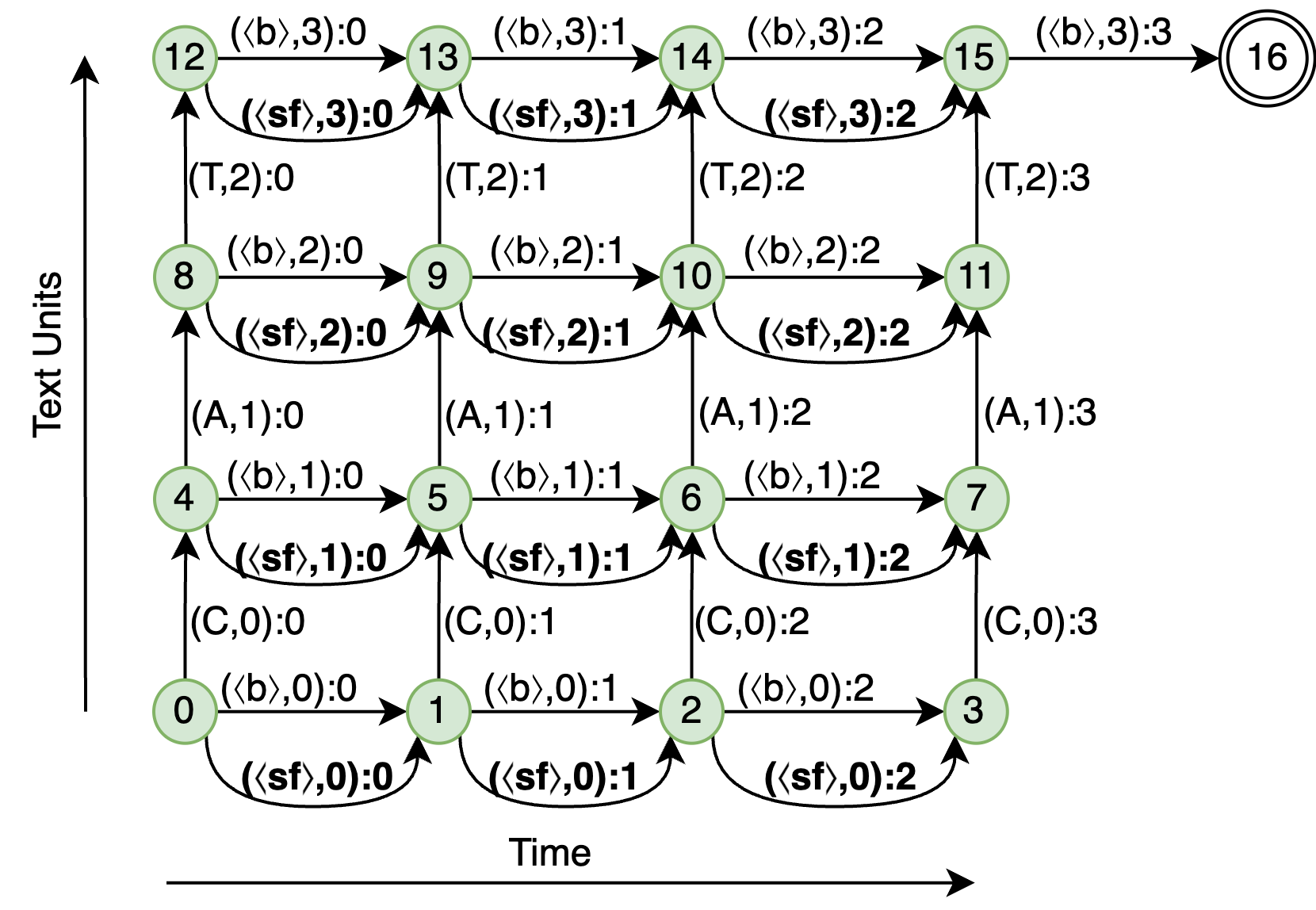}
        \label{fig:startlattice}
    }}
    \caption{WFSTs for Star-Transducer. $\langle sf \rangle$ is a special symbol indicating skipping the frame.}
    \label{fig:startschemas}
\end{figure}

We propose a simple but effective modification of the RNN-T loss computational graph to solve the problem of deletions. Star Transducer takes into account, along with the alignments with blank labels, the sequences when the blank label is substituted with a special "skip frame" $\langle sf \rangle$ symbol, which can be viewed as an allowance to skip frames produced by the encoder in training time. This approach is similar to the "*" token used in Star Temporal Classification~\cite{pratap2022stc} loss for CTC. For such frames, the transcription is missing in the ground truth, and the core idea was to allow skipping such frames when considering all possible alignments for loss computation. We add parallel arcs to those with $\langle b \rangle$ label to achieve this, as shown in~\ref{fig:startlattice}. Unlike other arcs, the weight for this arc is a hyperparameter and assigned directly after populating the lattice with other weights.

The Star-Transducer loss is implemented in the \href{https://github.com/artbataev/uol_final/blob/main/min_rnnt/losses/star_transducer.py}{\texttt{GraphStarTransducerLoss}} class.

\subsubsection{Bypass-Transducer (Bypass-T)}
\label{sec:bypass-transducer-explained}

\begin{figure}[ht]
    \centering
    \subfloat[\centering Bypass-Transducer Unit Schema. Labels: $(text\_unit,unit\_position):text\_unit$]{{
        \includegraphics[width=7.5cm]{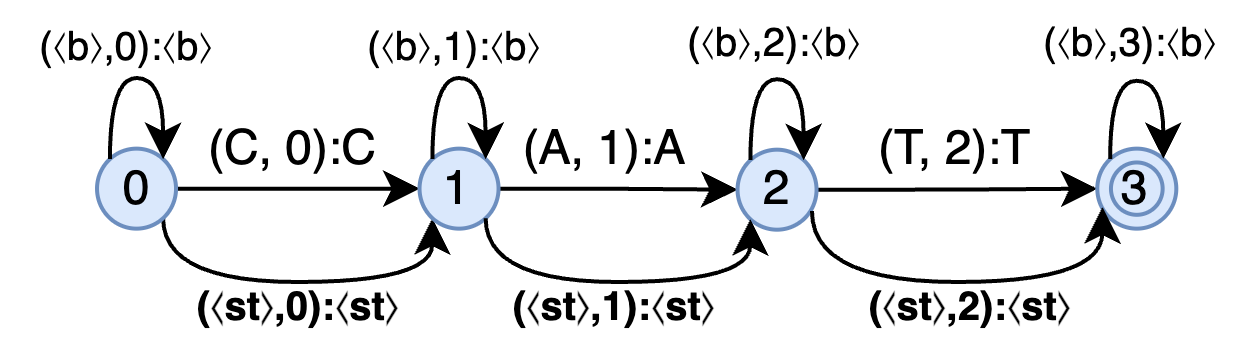}
        \label{fig:bypasstunit}
    }}
    \subfloat[\centering Bypass-Transducer Time Schema. Labels: $text\_unit:frame\_number$]{{
        \includegraphics[width=7.5cm]{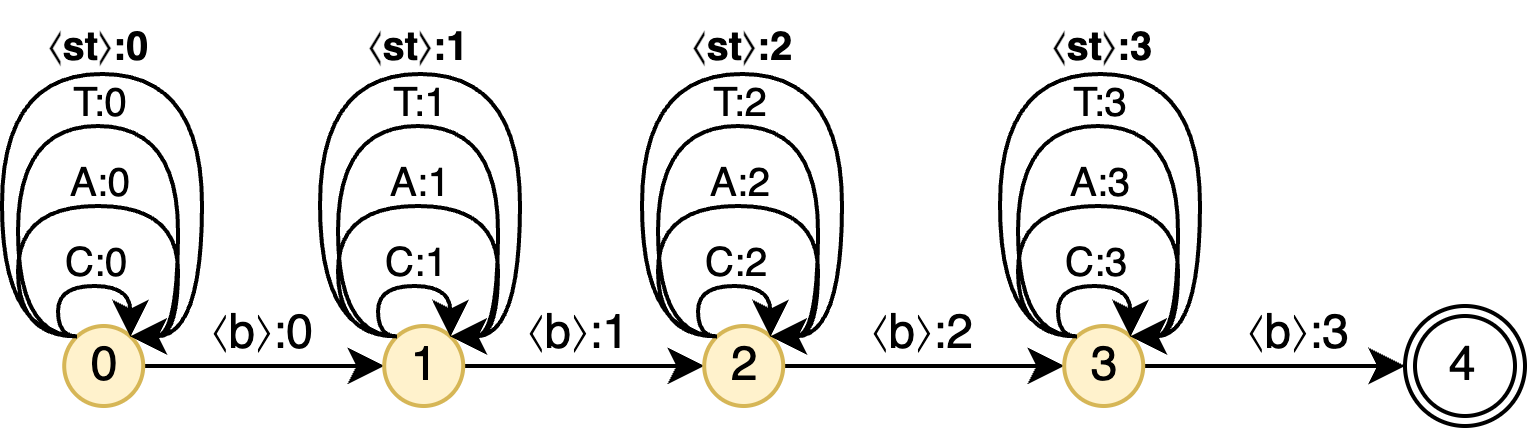}
        \label{fig:bypassttime}
    }}
    \qquad
    \subfloat[\centering Bypass-Transducer Lattice. Labels: $(text\_unit,unit\_position):frame\_number$]{{
        \includegraphics[width=12cm]{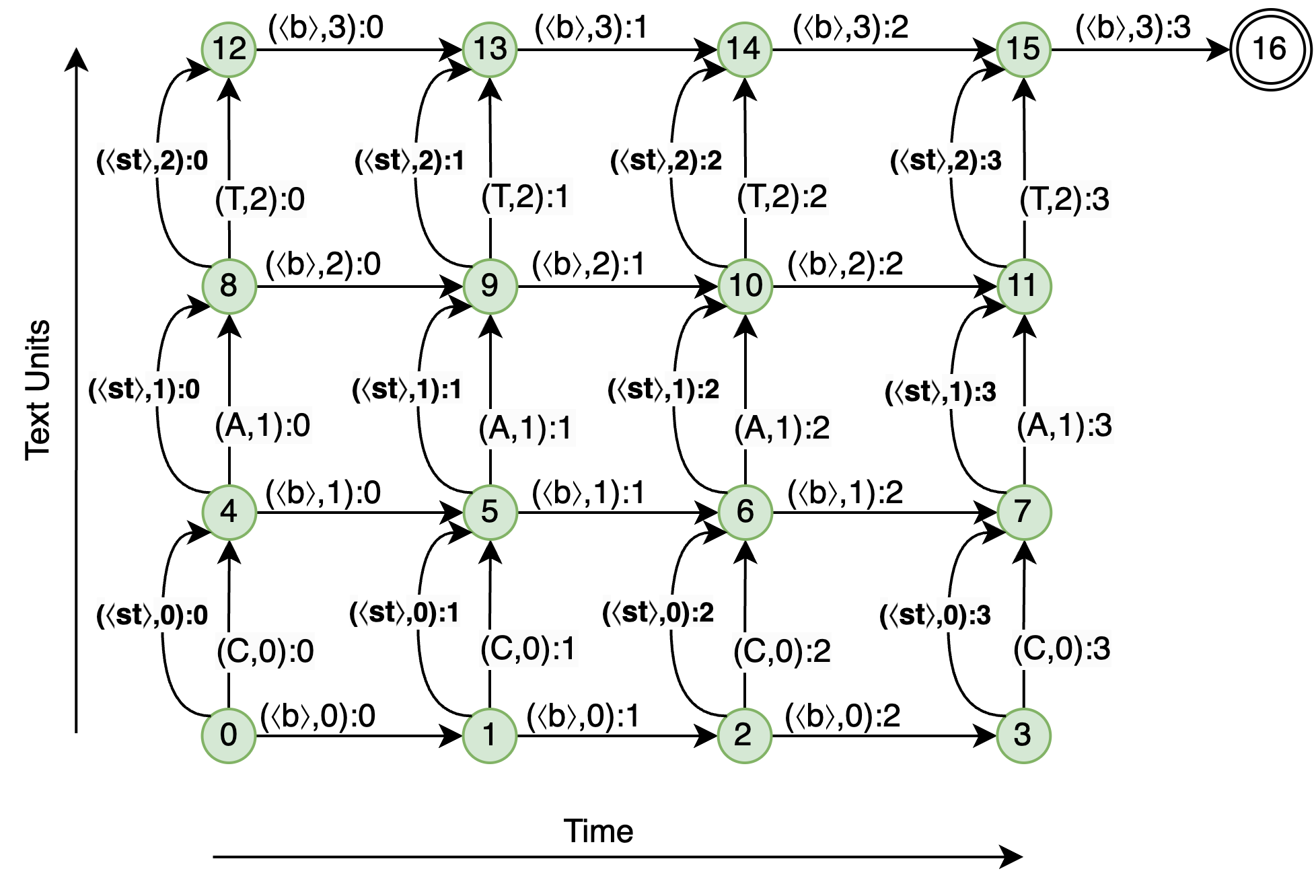}
        \label{fig:bypasstlattice}
    }}
    \caption{WFSTs for Bypass-Transducer. $\langle st \rangle$ is a special symbol indicating skipping the token.}
    \label{fig:bypasstschemas}
\end{figure}

For dealing with insertions, we propose a modification of the RNN-T computational graph, adding arcs with a special "skip token" $\langle st \rangle$ symbol, inspired by Bypass Temporal Classification~\cite{gao2023bypassbtc} approach. These arcs are parallel to the arcs with tokens. This means that the loss can consider alignments where some tokens are skipped. Fig.~\ref{fig:bypasstschemas} shows the temporal and unit schemas and the full constructed lattice.

The Bypass-Transducer loss is implemented in the \href{https://github.com/artbataev/uol_final/blob/main/min_rnnt/losses/bypass_transducer.py}{\texttt{GraphBypassTransducerLoss}} class.

In our experiments, we found that assigning constant weight similar to the Star-Transducer approach does not work. With small absolute values (e.g, $0$ or $-3$) the model is prone to produce deletions, and the system behaves worse than the original RNN-T. With high absolute weight values (e.g., $-20$), such transitions do not contribute to the loss computation: $e^{-20}$ is close to zero, and the loss is close to the original RNN-T. Similar to the approaches applied in the paper about BTC~\cite{gao2023bypassbtc}, we apply a schedule to the penalty weight ($skip\_token\_penalty$), combined with the probability derived from the output of the Joint network. For the probability we considered different options ($skip\_token\_mode$ parameter in the implementation):
\begin{itemize}
    \item "constant": only penalty constant, similar to one used in the Star-Transducer loss.
    \item "mean": mean probability for all labels (in log scale) excluding blank, similar to BTC~\cite{gao2023bypassbtc}.
    \item "max": maximum log-probability for all labels excluding blank.
    \item "maxexcl": maximum of the log probabilities of all labels excluding blank and ground truth labels.
    \item "sumexcl": logarithm of the sum of the probabilities (in log scale) of all labels excluding blank and ground truth labels.
\end{itemize}

We found that the "mean" and "max" options were not better than the original RNN-T loss. 
The "maxexcl" option was the first working solution used in the Preliminary Report. 
The intuition behind the "sumexcl" option is to assign the "unused" probability of outputs. The "sumexcl" option allows for the alignments to be considered when the network outputs a high probability for any token other than the target as "appropriate." We found that the "sumexcl" option outperforms other cases, as discussed further in Section~\ref{sec:bypass-transducer-eval}.

\subsubsection{Target-Robust-Transducer (TRT)}
\label{sec:target-robust-transducer-explained}

\begin{figure}[ht]
    \centering
    \subfloat[\centering Target-Robust-Transducer Unit Schema. Labels: $(text\_unit,unit\_position):text\_unit$]{{
        \includegraphics[width=7.5cm]{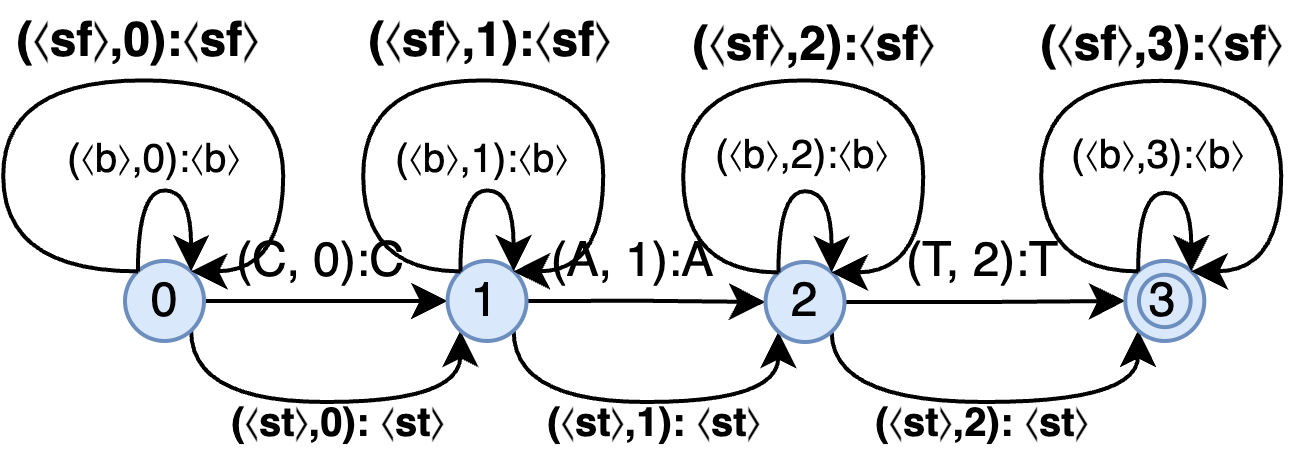}
        \label{fig:trtunit}
    }}
    \subfloat[\centering Target-Robust-Transducer Time Schema. Labels: $text\_unit:frame\_number$]{{
        \includegraphics[width=7.5cm]{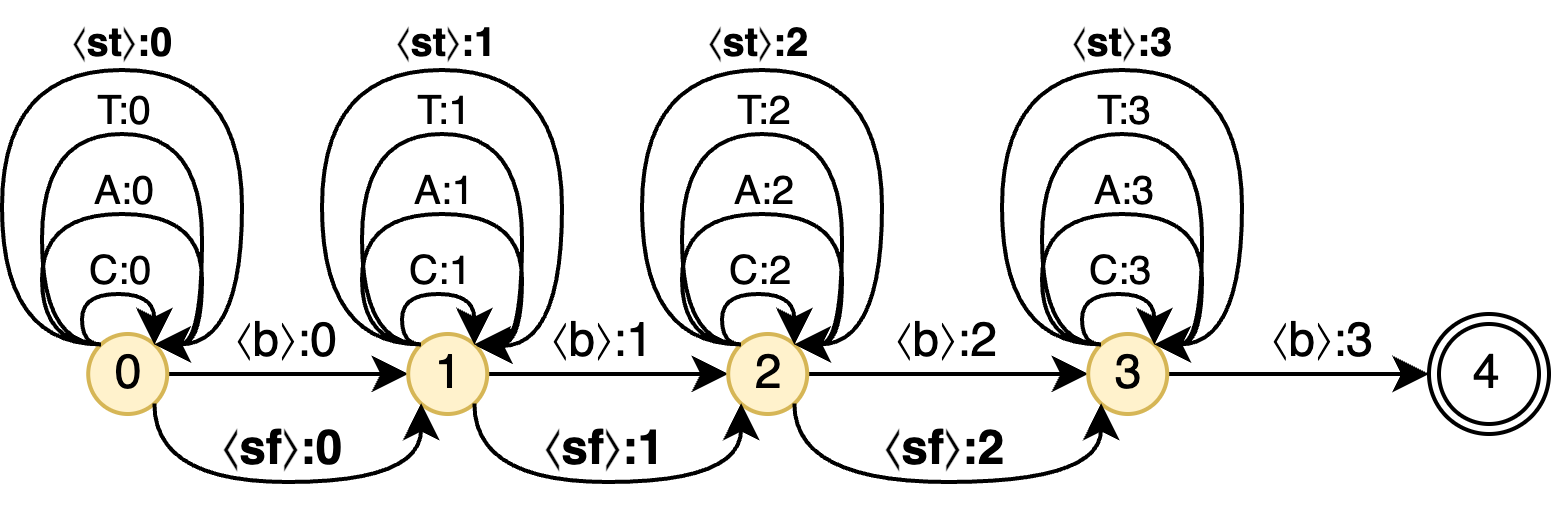}
        \label{fig:trttime}
    }}
    \qquad
    \subfloat[\centering Target-Robust-Transducer Lattice. Labels: $(text\_unit,unit\_position):frame\_number$]{{
        \includegraphics[width=12cm]{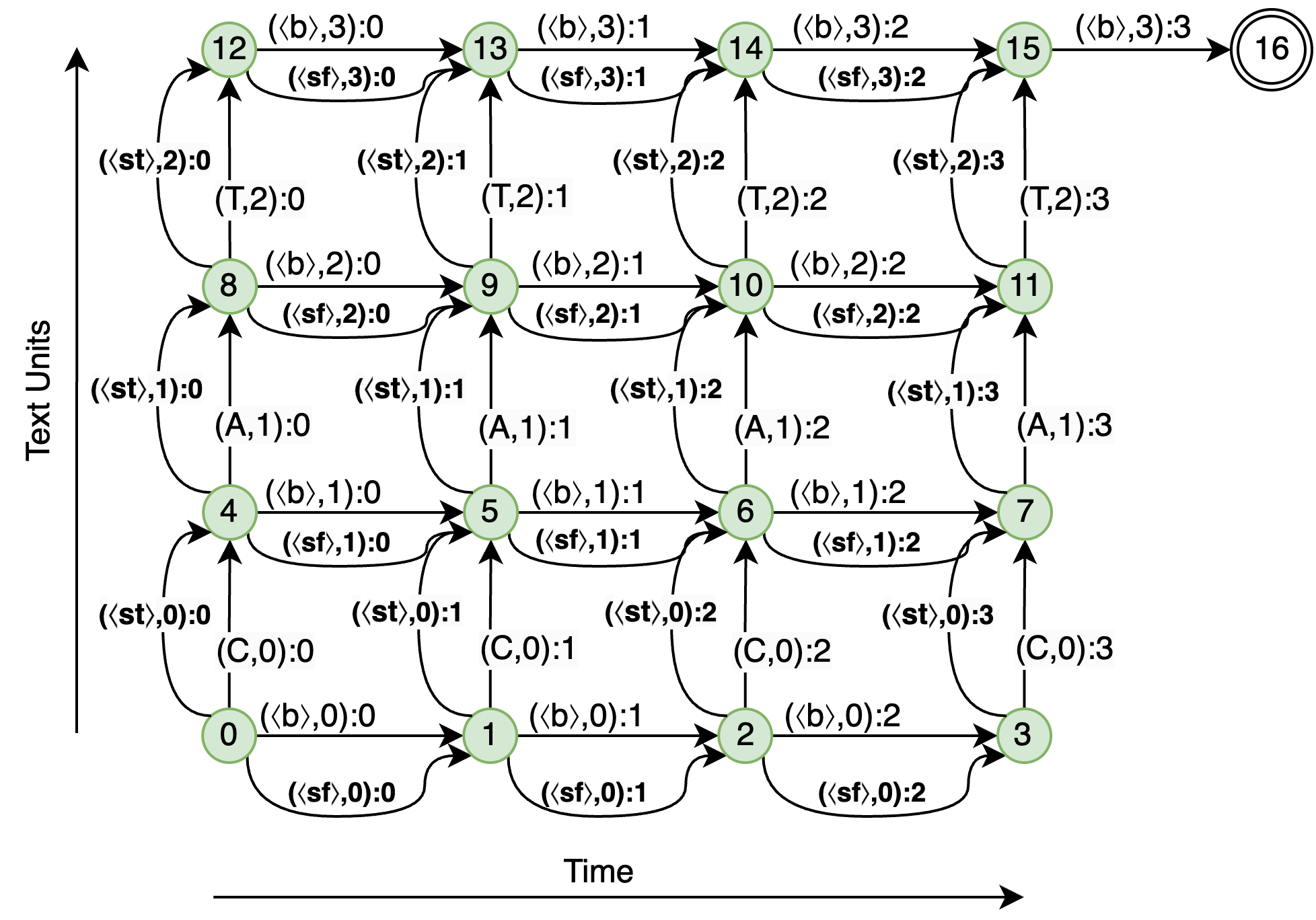}
        \label{fig:trtlattice}
    }}
    \caption{WFSTs for Target-Robust-Transducer. $\langle sf \rangle$ is a special symbol indicating skipping the frame. $\langle st \rangle$ is a special symbol indicating skipping the token.}
    \label{fig:trtschemas}
\end{figure}

Target-Robust-Transducer loss is a combination of Star-Transducer and Bypass-Transducer. We add both types of arcs that allow skipping frames and tokens, as shown in Fig.~\ref{fig:trtschemas}. 
It is worth mentioning that assigning $-\infty$ weight for "skip frame" arcs makes the loss identical to Bypass-Transducer (skipping frames is not allowed in this case), and $-\infty$ weight for "skip token" arcs makes it similar to Star-Transducer (skipping tokens is not allowed). This makes this loss a universal replacement for the previous two modifications (but the system makes more computations since the arcs are still present, even with $-\infty$ weight). We also test this behavior in unit tests.

The Target-Robust-Transducer loss is implemented in the class \\ \href{https://github.com/artbataev/uol_final/blob/main/min_rnnt/losses/target_robust_transducer.py}{\texttt{GraphTargetRobustTransducerLoss}}. The implementation combines hyperparameters and code for \texttt{GraphStarTransducerLoss} and \texttt{GraphBypassTransducerLoss}.

\section{Evaluation}

\subsection{Baseline}

\begin{table}[t!]
\caption{\label{tab:baseline} Baseline on LibriSpeech, WER [\%].}
\begin{center}
\begin{tabular}{ l | c c c c}
 \multirow{2}{*}{\textbf{Source}} & \multicolumn{2}{c|}{\textbf{dev}} & \multicolumn{2}{c|}{\textbf{test}} \\
 & \textbf{clean} & \multicolumn{1}{c|}{\textbf{other}} & \textbf{clean} & \multicolumn{1}{c|}{\textbf{other}} \\
 \toprule
 NeMo & 2.0 & 5.0 & 2.2 & 5.0\\
 Ours (200 epochs) & 2.1 & 4.9 & 2.2 & 5.1 \\
 \midrule
 Ours (60 epochs) & 2.6 & 6.8 & 2.8 & 6.8 \\
 Ours (100 epochs) & 2.4 & 5.9 & 2.5 & 6.0 \\
 \end{tabular}
\end{center}
\end{table}

The results for our implementation are shown in Table~\ref{tab:baseline}. For comparison, we use a publicly available Fast Conformer checkpoint\footnote{\begin{scriptsize}\url{https://catalog.ngc.nvidia.com/orgs/nvidia/teams/nemo/models/stt_en_fastconformer_transducer_large_ls}\end{scriptsize}} trained on LibriSpeech data for 200 epochs. Our implementation provides results comparable to those of the state-of-the-art pipeline. So we can proceed further and investigate the system behavior of corrupted targets. Additionally, we show the results for 60 and 100 epochs (also using the best checkpoint selected on \texttt{dev-other} for these epochs): to save computational resources, we evaluate different cases, training the models for 60 epochs, and for the final case with arbitrary errors we train the system for 100 epochs.

\subsection{Error Impact Exploration}

\begin{table}[t!]
\caption{\label{tab:corrupt-impact} Training RNN-T on data with errors, Fast Conformer, 60 epochs, WER [\%].}
\begin{center}
\begin{tabular}{ l | l | c c c c | c}
 \multirow{2}{*}{\textbf{Type}} & \multicolumn{1}{c|}{\multirow{2}{*}{\textbf{Corrupt \%}}} & \multicolumn{2}{c|}{\textbf{dev}} & \multicolumn{2}{c|}{\textbf{test}} & \textbf{WERD$\downarrow$} \\
 & & \textbf{clean} & \multicolumn{1}{c|}{\textbf{other}} & \textbf{clean} & \textbf{other} & \\
 \toprule
 \multicolumn{2}{c|}{–} & 2.6 & 6.8 & 2.8 & 6.8 & \\
 \midrule
 DEL & 20\% & 4.3 & 9.9 & 4.7 & 10.3 & 3.5 \\
 DEL & 50\% & 79.2 & 81.7 & 80.3 & 81.4 & 74.6 \\
 \midrule
 SUB & 20\% & 4.0 & 9.4 & 3.9 & 9.7 & 2.9 \\
 SUB & 50\% & 11.5 & 23.2 & 11.2 & 23.8 & 17.0 \\
 \midrule
 INS & 20\% & 4.0 & 10.3 & 4.2 & 10.2 & 3.4 \\
 INS & 50\% & 5.1 & 12.7 & 5.3 & 13.5 & 6.7 \\
\end{tabular}
\end{center}
\end{table}

To explore the training pipeline on partially incorrect transcripts, we generate additional training data sets by mutating the original training texts with the mutation probability $p_m$ of 20\% and 50\%. We are randomly removing words for the "deletions" case. We use randomly selected words from the training vocabulary for substitutions and insertions, substituting/inserting words with the probability $p_m$. 

Table~\ref{tab:corrupt-impact} shows the training results on corrupted transcripts. With a small amount of corruption, all cases lead to system degradation, but the difference between cases is tiny (from 2.9\% to 3.5\% absolute WER degradation on test-other). We found that the deletions are most disruptive for the high corruption rate of 50\%, and the ASR system can not achieve a reasonable quality  (81.4\% WER on test-other compared to 6.8\% on original data). Thus, we prioritized the work with this part of the problem.
Substitutions are the next hard case for RNN-T, which is the opposite of observations for the behavior of CTC systems in~\cite{gao2023omniotc}.

\subsection{Dealing with Deletions: Star Transducer}

\begin{table}[t!]
\caption{\label{tab:star-transducer} Star-Transducer (Star-T) Loss for deletions, Fast Conformer, 60 epochs, WER [\%].}
\begin{center}
\begin{tabular}{ l | l | l | c c c c | c | c}
 \multirow{2}{*}{\textbf{Loss}} & \textbf{Skip} & \multirow{2}{*}{\textbf{DEL \%}} & \multicolumn{2}{c|}{\textbf{dev}} & \multicolumn{2}{c|}{\textbf{test}} & \textbf{WERD$\downarrow$} & \textbf{WERDR$\uparrow$} \\
 & \textbf{Weight} &  & \textbf{clean} & \multicolumn{1}{c|}{\textbf{other}} & \textbf{clean} & \textbf{other} & & \\
 \toprule
 RNN-T & - & – & 2.6 & 6.8 & 2.8 & 6.8 & &\\
 \midrule
 RNN-T & - & 20\% & 4.3 & 9.9 & 4.7 & 10.3 & 3.5 & \\
 \textbf{Star-T} & 0 & 20\% & 3.9 & 8.2 & 4.3 & 8.5 & 1.7 & 51.4\%\\
 \textbf{Star-T} & -0.5 & 20\% & 3.1 & 7.5 & 3.4 & \textbf{7.6} & 0.8 & \textbf{77.1\%} \\
 \midrule
 RNN-T & - & 50\% & 79.2 & 81.7 & 80.3 & 81.4 & 74.6 & \\
 \textbf{Star-T} & 0 & 50\% & 5.1 & 10.6 & 5.2 & \textbf{11.0} & 4.2 & \textbf{94.4\%} \\
 \textbf{Star-T} & -0.5 & 50\% & 5.4 & 12.4 & 5.9 & 12.5 & 5.7 & 92.4\% \\
\end{tabular}
\end{center}
\end{table}

For the setup when the ground truth transcripts contain deletions, we apply Star-Transducer loss as a drop-in replacement for the RNN-T loss.
The results of training the model are shown in Table~\ref{tab:star-transducer}. In both scenarios, we can close the gap between the baseline for more than 70\%:  77.1\% WERDR for 20\% deletions and 94.4\% for 50\%. We found that the training is stable even without penalty, but applying the small constant penalty for the "skip frame" transition ($-0.5$) improves the quality when the number of deletions is low. 
We were surprised that such a simple solution works and that modifying the autoregressive prediction network is unnecessary. This can mean that the encoder and joint are more sensitive to incorrect transcripts than the prediction network.

\subsection{Dealing with Insertions: Bypass Transducer}
\label{sec:bypass-transducer-eval}

\begin{table}[t!]
\caption{\label{tab:bypass-transducer} Bypass-Transducer (Bypass-T) Loss for insertions, Fast Conformer, 60 epochs, WER [\%].}
\begin{center}
\begin{tabular}{ l | l | l | c c c c | c | c}
 \multirow{2}{*}{\textbf{Loss}} & \multicolumn{1}{c|}{\textbf{Skip}} & \multirow{2}{*}{\textbf{INS}} & \multicolumn{2}{c|}{\textbf{dev}} & \multicolumn{2}{c|}{\textbf{test}} & \textbf{WERD$\downarrow$} & \textbf{WERDR$\uparrow$} \\
 & \textbf{Weight} &  & \textbf{clean} & \multicolumn{1}{c|}{\textbf{other}} & \textbf{clean} & \textbf{other} & & \\
 \toprule
 RNN-T & - & – & 2.6 & 6.8 & 2.8 & 6.8 & &\\
 \midrule
 RNN-T & – & 20\% & 4.0 & 10.3 & 4.2 & 10.2 & 3.4 & \\
 \textbf{Bypass-T} & -6 & 20\% & 3.0 & 7.5 & 3.3 & 7.9 & 1.1 & \textbf{67.6\%} \\
 \midrule
 RNN-T & – & 50\% & 5.1 & 12.7 & 5.3 & 13.5 & 6.7 & \\
 \textbf{Bypass-T} & -6 & 50\% & 3.9 & 10.3 & 4.3 & 10.5 & 3.7 & 44.8\% \\
 \textbf{Bypass-T} & -5 & 50\% & 3.6 & 9.2 & 4.0 & 9.4 & 2.6 & \textbf{61.2\%} \\
\end{tabular}
\end{center}
\end{table}

For insertions case, we apply the Bypass-Transducer loss described in Section~\ref{sec:bypass-transducer-explained}. 
The results are shown in the Table~\ref{tab:bypass-transducer}. The transition weight for the "skip token" arcs is a sum of the constant weight and the total log-probability of all outputs excluding blank and target ("sumexcl" option), as discussed in the Section~\ref{sec:bypass-transducer-explained}.
The training starts with a constant weight of $-20.0$ and is adjusted with the decay after each epoch (starting 3rd epoch): $weight_{next} = min(max\_weight, weight * decay)$. We use $decay = 0.9$ for all experiments. Table~\ref {tab:bypass-transducer} also reports the maximum constant penalty applied in training.
The proposed loss can restore more than 60\% of the system quality for the texts with insertions. Further evaluation of the "sumexcl" and "maxexcl" options for assigning the weight can be found in Appendix~\ref{sec:bypass-transducer-eval-extended}.

\subsection{Dealing with Substitutions: Target-Robust-Transducer}

\begin{table}[t!]
\caption{\label{tab:target-robust-transducer-sub} Target Robust Transducer (TRT) Loss for substitutions, Fast Conformer, 60 epochs, WER [\%].}
\begin{center}
\begin{tabular}{ l | l | l | c c c c | c | c}
 \multirow{2}{*}{\textbf{Loss}} & \textbf{Skip} & \multirow{2}{*}{\textbf{SUB}} & \multicolumn{2}{c|}{\textbf{dev}} & \multicolumn{2}{c|}{\textbf{test}} & \textbf{WERD$\downarrow$} & \textbf{WERDR$\uparrow$} \\
 & token,frame &  & \textbf{clean} & \multicolumn{1}{c|}{\textbf{other}} & \textbf{clean} & \textbf{other} & & \\
 \toprule
 RNN-T & - & – & 2.6 & 6.8 & 2.8 & 6.8 & &\\
 \midrule
 RNN-T & - & 20\% & 4.0 & 9.4 & 3.9 & 9.7 & 2.9 & \\
 TRT & -8,-0.5 & 20\% & 3.4 & 8.2 & 3.8 & 8.5 & 1.7 & \textbf{41.4\%}\\
 \midrule
 RNN-T & - & 50\% & 11.5 & 23.2 & 11.2 & 23.8 & 17.0 & \\
 \textbf{TRT} & -8,-0.5 & 50\% & 8.2 & 16.3 & 8.5 & 17.0 & 10.2 & 45.3\% \\
 \textbf{TRT} & -8,-1 & 50\% & 6.9 & 16.0 & 7.1 & 15.8 & 9.0 & \textbf{47.1\%} \\
\end{tabular}
\end{center}
\end{table}

We apply Target-Robust-Transducer for the case with substitutions since any "substitution" can be viewed as a combination of "deletion" and "insertion." The results are shown in Table~\ref{tab:target-robust-transducer-sub}. In the preliminary experiments, we found that assigning low absolute values for weights ($0$ for skip frame as in Star-Transducer, and $-5$ or $-6$ for skip token as in Bypass-Transducer) results in fast model overfitting, but when the penalty is more significant, the training is stable. Since the loss can skip both frames and tokens, applying a more significant penalty is reasonable.
We use $-8$ for skip frame penalty and the "sumexcl" option for assigning the weight, which we found the best when experimenting with Bypass-Transducer, along with the penalty schedule, as discussed above. With the proposed loss, we can restore more than 40\% of the system degradation on \texttt{test-other}: we achieve WERDR of 41.4\% for 20\% substitutions and 47.1\% for 50\%.

\subsection{Arbitrary Errors}
\label{sec:trt-arbitrary-errors}

\begin{table}[t!]
\caption{\label{tab:target-robust-transducer-all} Target Robust Transducer (TRT) Loss for arbitrary errors, Fast Conformer, 60 and 100 epochs, WER [\%]. 50\% of data is corrupted, using 15\% for each class of errors.}
\begin{center}
\begin{tabular}{ l | l | l | c c c c | c | c}
 \multirow{2}{*}{\textbf{Loss}} & \multirow{2}{*}{\textbf{Epochs}} & \multirow{2}{*}{\textbf{ERR}} & \multicolumn{2}{c|}{\textbf{dev}} & \multicolumn{2}{c|}{\textbf{test}} & \textbf{WERD$\downarrow$} & \textbf{WERDR$\uparrow$} \\
 &  &  & \textbf{clean} & \multicolumn{1}{c|}{\textbf{other}} & \textbf{clean} & \textbf{other} & & \\
 \toprule
 RNN-T & 60 & - & 2.6 & 6.8 & 2.8 & 6.8 & &\\   
 \midrule
 RNN-T & 60 & 50\% & 4.2 & 10.2 & 4.3 & 10.1 & 3.3 & \\
 \textbf{TRT} & 60 & 50\% & 3.3 & 8.0 & 3.6 & 8.4 & 1.6 & \textbf{51.5}\% \\
 \midrule
 \midrule
 RNN-T & 100 & – & 2.4 & 5.9 & 2.5 & 6.0 & &\\
 \midrule
 RNN-T & 100 & 50\% & 3.5 & 9.4 & 3.8 & 9.5 & 3.5 & \\
 \textbf{TRT} & 100 & 50\% & 2.9 & 7.0 & 3.2 & 7.0 & 1.0 & \textbf{71.4}\% \\
\end{tabular}
\end{center}
\end{table}

For evaluating the system trained with Target-Robust-Transducer loss, we construct the extra training data by corrupting only 50\% of all utterances. For each corrupted utterance, we apply random substitutions, insertions, and deletions with probability for each type of 15\%. We consider such case closer to the actual conditions used for production systems when the well-curated datasets are mixed with unreliable data from different sources. Also, we train the system for longer (100 epochs). As shown in Table~\ref{tab:target-robust-transducer-all}, we are able to restore more than 51\% system quality when the system is trained for 60 epochs (compared to RNN-T baseline also trained for 60 epochs). When trained for an extra 40 epochs, the system can restore more than 71\% quality (WERDR 71.4\%).
In the Appendix~\ref{sec:appendix-arb-err-trt}, we also publish the learning curves for the system demonstrating its effectiveness in reducing substitutions and deletions on the \texttt{dev-clean} data. 


\section{Conclusion}

In our project, we trained speech recognition neural systems on the LibriSpeech~\cite{panayotov2015librispeech} dataset. We explored the system's robustness to errors in target texts by artificially corrupting the ground truth target texts from the dataset. We also explored different RNN-T loss modifications to solve the problem of quality degradation in the discussed scenarios and proposed three losses: 
\begin{itemize}
    \item \textbf{Star-Transducer}, which mitigates the effect of missing words in transcripts and is able to restore more than 90\% of the system quality in such case
    \item \textbf{Bypass-Transducer}, which allows insertions (extra words) in the transcripts and allows the restoration of more than 60\% of the quality in such cases compared to "clean" transcripts
    \item \textbf{Target-Robust-Transducer}, which combines the approaches applied in the previous two losses. This loss can deal with arbitrary types of errors. It improves the system's quality when some words of transcripts are incorrect (substitutions), mitigating more than 40\% of the quality loss for this case. For arbitrary types of errors, we also show that it can restore more than 70\% of the quality compared to the baseline with the well-transcribed data.
\end{itemize}

Our work is based on the previous solutions for CTC loss~\cite{pratap2022stc,gao2023bypassbtc,gao2023omniotc} and Graph-RNN-T framework~\cite{laptev2023rnntwfst}, and proposes a novel valuable solution for RNN-Transducer-based ASR systems.
We demonstrated the effectiveness of the losses using the Fast Conformer~\cite{rekesh2023fastconformer} model.

The proposed Target-Robust-Transducer system can be applied in real-world scenarios when training models on a large amount of data from unreliable sources that usually contain transcription errors.

We also see direct applications for Star-Transducer beyond the discussed case with missing words in transcripts. Modern ASR systems are trained not only to provide transcription (in words) but also to provide punctuation, e.g., Whisper~\cite{radford2023whisper}. Since many curated ASR corpora do not contain punctuation (e.g., LibriSpeech~\cite{panayotov2015librispeech} which we use in our work), such missing punctuation can be viewed as "deletions" in the transcripts, and with Star-Transducer loss the model can be trained directly on a mixture of datasets with and without punctuation.

In further work, we plan to investigate other models (e.g., Conformer~\cite{gulati2020conformer} with 4x subsampling) and datasets. We also plan to apply the losses to train ASR systems on a large scale for production usage.

The losses are planned to be proposed to the open-source NeMo~\cite{kuchaiev2019nemo} framework.


\section{Report Parameters and Additional Notes}

The implementation is published in the GitHub repository: \url{https://github.com/artbataev/uol_final}.

The report contains 6 tables and 6 figures. The appendix contains an additional 1 table and 1 figure.
We comply with word limits for each section.

\section{Acknowledgments}
I want to express my gratitude to my employer, NVIDIA Corporation, for providing computational resources for this research.
\addcontentsline{toc}{section}{References}
\bibliographystyle{IEEEtran}
\bibliography{references}
\clearpage

\begin{appendices}

\section{Arbitrary Errors - Learning Curve}
\label{sec:appendix-arb-err-trt}

\begin{figure}[h]
\centering
\includegraphics[width=16cm]{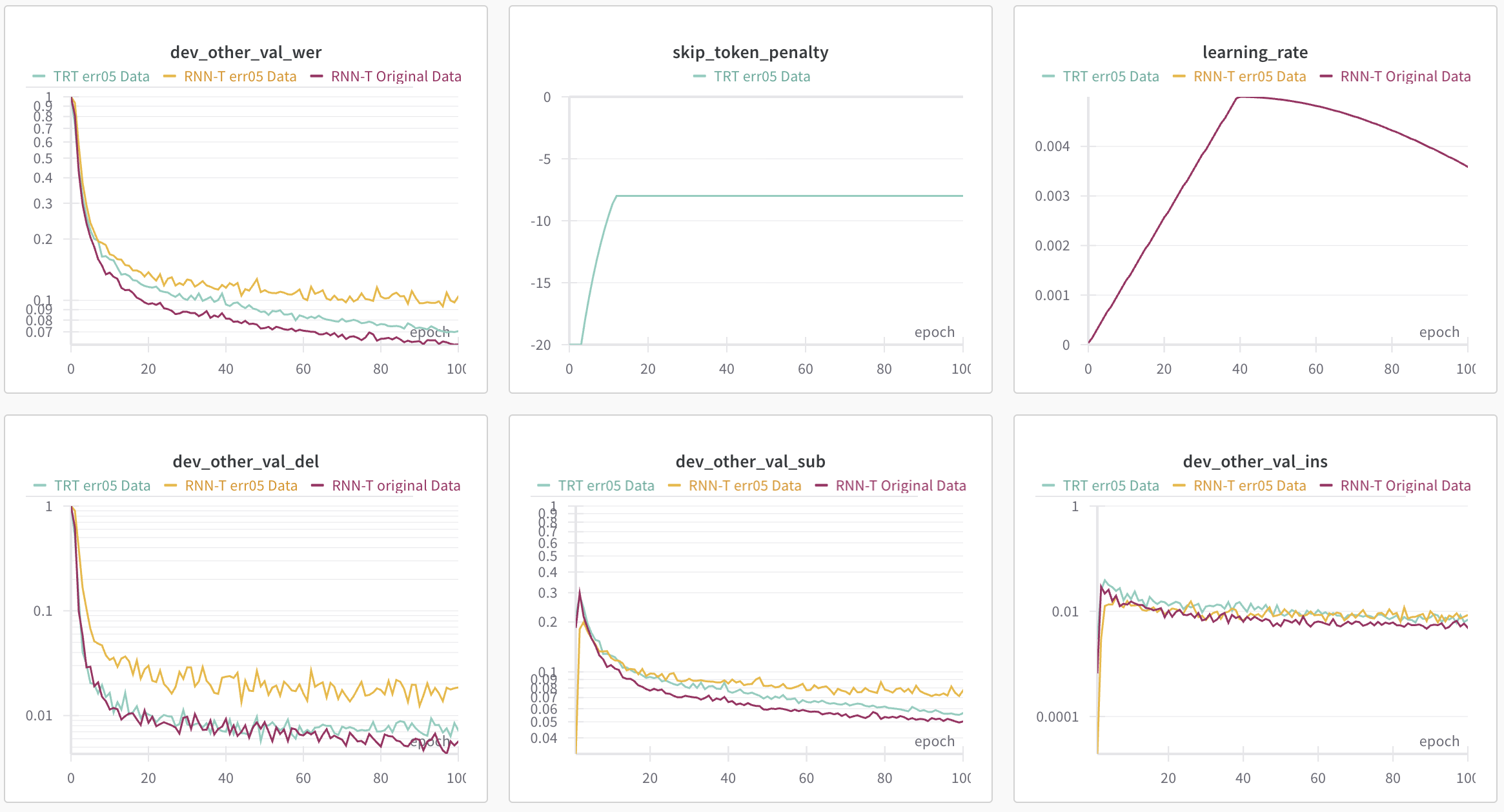}
\caption{Arbitrary errors: learning curves for RNN-T (original and corrupted data) and Target Robust Transducer (corrupted data).}
\label{fig:trtplot}
\end{figure}

We provide an additional plot with the learning curve, demonstrating WER and its components for RNN-T and Target Robust Transducer training with arbitrary errors for 100 epochs, as discussed in Section~\ref{sec:trt-arbitrary-errors}. We can see in Figure~\ref{fig:trtplot} that the number of insertions produced in all cases is similar, but the number of deletions and substitutions produced by the system trained on corrupted data with the TRT loss is significantly lower than for RNN-T and is close to the number of errors produced by the RNN-T on the original non-corrupted data. The screenshot is produced by WandB~\cite{wandb}.

\section{Bypass-Transducer: Extended evaluation of hyperparameters}
\label{sec:bypass-transducer-eval-extended}

In this appendix section, we show the extended hyperparameter evaluation of the options for Bypass-Transducer loss regarding assigning weights for skip token transitions.
In initial experiments, as discussed in Section~\ref{sec:bypass-transducer-explained}, we tried different options and found that the system is trainable only with "maxexcl" and "sumexcl" options. In the system exploration process, we found the "sumexcl" option, which considers total "unassigned" log-probability (log-probability for all outputs excluding blank and target labels) to provide the best value. We provide an extended version of the Table~\ref{tab:bypass-transducer}. The results in~\ref{tab:bypass-transducer-extended} show that the "sumexcl" option outperforms the "maxexcl" by a significant margin.

\begin{table}[t!]
\caption{\label{tab:bypass-transducer-extended} Bypass-Transducer Loss for insertions, Fast Conformer, 60 epochs, WER [\%]. Extended evaluation.}
\begin{center}
\begin{tabular}{ l | l | l | c c c c | c | c}
 \multirow{2}{*}{\textbf{Loss}} & \multicolumn{1}{c|}{\textbf{Skip}} & \multirow{2}{*}{\textbf{INS}} & \multicolumn{2}{c|}{\textbf{dev}} & \multicolumn{2}{c|}{\textbf{test}} & \textbf{WERD$\downarrow$} & \textbf{WERDR$\uparrow$} \\
 & \textbf{Weight,Mode} &  & \textbf{clean} & \multicolumn{1}{c|}{\textbf{other}} & \textbf{clean} & \textbf{other} & & \\
 \toprule
 RNN-T & - & – & 2.6 & 6.8 & 2.8 & 6.8 & &\\
 \midrule
 RNN-T & – & 20\% & 4.0 & 10.3 & 4.2 & 10.2 & 3.4 & \\
 \textbf{Bypass-T} & -6,maxexcl & 20\% & 3.2 & 7.9 & 3.3 & 8.0 & 1.2 & 65.7\% \\
 \textbf{Bypass-T} & -6,sumexcl & 20\% & 3.0 & 7.5 & 3.3 & 7.9 & 1.1 & \textbf{67.6\%} \\
 \midrule
 RNN-T & – & 50\% & 5.1 & 12.7 & 5.3 & 13.5 & 6.7 & \\
 \textbf{Bypass-T} & -6,maxexcl & 50\% & 4.4 & 10.3 & 4.1 & 10.7 & 3.9 & 41.8\% \\
 \textbf{Bypass-T} & -6,sumexcl & 50\% & 3.9 & 10.3 & 4.3 & 10.5 & 3.7 & 44.8\% \\
 \textbf{Bypass-T} & -5,maxexcl & 50\% & 4.1 & 10.2 & 4.5 & 10.4 & 3.6 & 46.3\% \\
 \textbf{Bypass-T} & -5,sumexcl & 50\% & 3.6 & 9.2 & 4.0 & 9.4 & 2.6 & \textbf{61.2\%} \\
\end{tabular}
\end{center}
\end{table}

\section{Losses Visualization}
\label{sec:appendix-visualization}
We additionally publish the Jupyter Notebook, which visualizes the lattices of the proposed losses. The notebook can be found in the repository \\ 
\url{https://github.com/artbataev/uol_final/blob/main/notebooks/Loss_Demo.ipynb}.

\end{appendices}


\end{document}